\documentclass[11pt]{article}

\usepackage{scicite}

\usepackage{times}

\usepackage{tablefootnote}
\usepackage{multicol}

\usepackage{amsmath}
\usepackage{graphicx,float,psfrag,epsf,subcaption}
\usepackage[labelfont=bf]{caption}
\usepackage{enumerate}
\usepackage{array}
\usepackage{tablefootnote}
\usepackage{longtable}
\usepackage{setspace}
\usepackage{verbatim}
\usepackage{xcolor}
\usepackage{bbm}
\usepackage{algorithm,algorithmic}

\newenvironment{sciabstract}{%
\begin{quote} \bf}
{\end{quote}}

\newcounter{lastnote}

\title{Spatial Heterogeneity Can Lead to Substantial Local Variations in COVID-19 Timing and Severity}

\author
{Loring J. Thomas,$^{1}$ Peng Huang$^{1}$, Fan Yin$^{2}$, Xiaoshuang Iris Luo$^{3}$, \\
Zack W. Almquist$^{4}$, John R. Hipp$^{3}$, Carter T. Butts$^{125\ast}$\\
\\
\normalsize{$^{1}$Department of Sociology, University of California, Irvine}\\
\normalsize{$^{2}$Department of Statistics, University of California, Irvine}\\
\normalsize{$^{3}$Department of Criminology, University of California, Irvine}\\
\normalsize{$^{4}$Department of Sociology, University of Washington}\\
\normalsize{$^{5}$Department of Computer Science and EECS, University of California, Irvine}\\
\normalsize{$^\ast$To whom correspondence should be addressed; E-mail:  buttsc@uci.edu.}
}

\date{}

\begin{document}




\maketitle

\newcommand{\ctb}[1]{{\color{red}[CTB: #1]}}
\newcommand{\zwa}[1]{{\color{blue}[ZWA: #1]}}
\newcommand{\ljt}[1]{{\color{purple}[LJT:#1]}}


\begin{sciabstract}

Standard epidemiological models for COVID-19 employ variants of compartment (SIR) models at local scales, implicitly assuming spatially uniform local mixing.  Here, we examine the effect of employing more geographically detailed diffusion models based on known spatial features of interpersonal networks, most particularly the presence of a long-tailed but monotone decline in the probability of interaction with distance, on disease diffusion.  Based on simulations of unrestricted COVID-19 diffusion in 19 U.S cities, we conclude that heterogeneity in population distribution can have large impacts on local pandemic timing and severity, even when aggregate behavior at larger scales mirrors a classic SIR-like pattern.  Impacts observed include severe local outbreaks with long lag time relative to the aggregate infection curve, and the presence of numerous areas whose disease trajectories correlate poorly with those of neighboring areas.  A simple catchment model for hospital demand illustrates potential implications for health care utilization, with substantial disparities in the timing and extremity of impacts even without distancing interventions.  Likewise, analysis of social exposure to others who are morbid or deceased shows considerable variation in how the epidemic can appear to individuals on the ground, potentially affecting risk assessment and compliance with mitigation measures.  These results demonstrate the potential for spatial network structure to generate highly non-uniform diffusion behavior even at the scale of cities, and suggest the importance of incorporating such structure when designing models to inform healthcare planning, predict community outcomes, or identify potential disparities.
\end{sciabstract}

\section*{Introduction}

Since its emergence at the end of 2019, the SARS-CoV-2 virus has spread rapidly to all portions of globe, infecting nearly five million people as of late May 2020 \cite{world2020coronavirus}. The disease caused by this virus, denoted COVID-19, generally manifests as a respiratory illness that is spread primarily via airborne droplets.  While most cases of COVID-19 are non-fatal, a significant fraction of those infected require extensive supportive care, and the mortality rate is substantially higher than more common infectious diseases such as seasonal influenza \cite{onder2020case}.  Even for survivors, infection can lead to long-term damage to the lungs and other organs, leading to long convalescence times and enhance risks of secondary complications \cite{jiang2020review,geng2020pathophysiological}.  By early March of 2020, COVID-19 outbreaks had appeared on almost every continent, including significant clusters within many cities \cite{world2020coronavirus1}. In the absence of an effective vaccine, public health measures to counteract the pandemic in developed nations have focused on social distancing measures that seek to slow diffusion sufficiently to avoid catastrophic failure of the healthcare delivery system.  Both the planning and public acceptance of such measures have been highly dependent upon the use of epidemiological models to probe the potential impact of distancing interventions, and to anticipate when such measures may be loosened with an acceptable level of public risk.  As such, the assumptions and behavior of COVID-19 diffusion models is of significant concern.

Currently dominant approaches to COVID-19 modeling \cite{jackson2020effects,zhang2020impact,pujari2020multi} are based on compartment models (often called \emph{SIR} models, after the conventional division of the population into \emph{susceptible}, \emph{infected}, and \emph{recovered} groups in the most basic implementations) that implicitly treat individuals within a population as geographically well-mixed.  While some such models include differential contact by demographic groups (e.g., age), and may treat states, counties, or occasionally cities as distinct units, those models presently in wide use do not incorporate spatial heterogeneity at local scales (e.g., within cities).  Past work, however, has shown evidence of substantial heterogeneity in social relationships at regional, urban, and sub-urban scales \cite{spiro2016persistence,smith2015relationship},  with these variations in social network structure impacting outcomes as diverse as regional identification \cite{almquist2015predicting}, disease spread \cite{riley2007large}, and crime rates \cite{hipp2013extrapolative}, as well as in both human and non-human networks \cite{leu2016environment}.  If individuals are not socially ``well-mixed'' at local scales, then it is plausible that diffusion of SARS-CoV-2 via interpersonal contacts will likewise depart from the uniform mixing characteristic of the SIR models.  Indeed, at least one computational study \cite{almquist2012point} using a fairly ``generic'' (non-COVID) diffusion process on realistic urban networks has showed considerable non-uniformity in diffusion times, suggesting that such effects could hypothetically be present.  However, it could also be hypothesized that such effects would be small perturbations to the broader infection curve captured by conventional compartment models, with little practical importance.  The question of whether these effects are likely 
to be present for COVID-19, and if so their strength and size, has to date remained open.  

In this paper, we examine the potential impact of local spatial heterogeneity on COVID-19, modeling the diffusion of SARS-CoV-2 in populations whose contacts are based on spatially plausible network structures.  We focus here on the urban context, examining nineteen different cities in the United States.  We simulate the population of each city in detail (i.e., at the individual level), simulating hypothetical outbreaks on the contact network in each city in the absence of measures such as social distancing.  Despite allowing the population to be well-mixed in all other respects (i.e., not imposing mixing constraints based on demographic or other characteristics), we find that spatial heterogeneity alone is sufficient to induce substantial departures from spatially homogeneous SIR behavior.  Among the phenomena observed are ``long lag'' outbreaks that appear in previously unharmed communities after the aggregate infection wave has largely subsided; frequently low correlations between infection timing in spatially adjacent communities; and distinct sub-patterns of outbreaks found in some urban areas that are uncorrelated with the broader infection pattern.  Gaps between infection peaks at the intra-urban level can be be large, e.g. on the order of weeks or months in extreme cases, even for communities that are within kilometers of each other.  Such heterogeneity is potentially consequential for the management of healthcare delivery services: as we show using a simple ``catchment'' model of hospital demand, local variations in infection timing can easily overload some hospitals while leaving others relatively empty (absent active reallocation of patients).  Likewise, we show that individuals' social exposures to others who are morbid or deceased vary greatly over the course of the pandemic, potentially leading to differences in risk assessment and bereavement burden for persons residing in different locations.  Differences in outbreak timing and severity may exacerbate health disparities (since e.g., surge capacity varies by community) and may even affect perception of and support for prophylactic behaviors among the population at large, with those in so-far untouched communities falsely assuming that the pandemic threat is either past or was exaggerated to begin with, or attributing natural variation in disease timing to the impact of health interventions.

We note at the outset that the models used here are intended to probe the hypothetical impact of spatial heterogeneity on COVID-19 diffusion within particular scenarios, rather than to produce high-accuracy predictions or forecasts.  For the latter applications, it is desirable to incorporate many additional features that are here simplified to facilitate insight into the phenomenon of central interest. In particular, we do not incorporate either demographic effects or social distancing, allowing us consider a setting that is as well-mixed as possible (and hence as close as possible to an idealized SIR model) with the exception of spatial heterogeneity. As we show, even this basic scenario is sufficient to produce large deviations from the SIR model.  Despite the simplicity of our models, we do note that the approach employed here could be integrated with other factors and calibrated to produce models intended for forecasting or similar applications. 

\section*{Methods}

COVID-19 is typically transmitted via direct contact with infected individuals, with the greatest risk occurring when an uninfected person is within approximately six feet of an infected person for an extended period of time.  Such interactions can be modeled as events within a social network, where individuals are tied to those whom they have a high hazard of intensive interaction. In prior work, this approach has been successfully employed for modeling infectious disesaes ranging from HIV \cite{morris2004network} and influenza \cite{viboud2006synchrony} to Zika \cite{li2019analysis} transmission.  To model networks of potential contacts at scale, we employ spatial network models \cite{butts2011spatial}, which are both computationally tractable and able to capture the effects of geography and population heterogeneity on network structure \cite{butts2012geographical}.  Such models have been successfully used to capture social phenomena ranging from neighborhood-level variation crime rates \cite{hipp2013extrapolative} and regional identification \cite{almquist2015predicting} to the flow of information among homeless persons \cite{almquist2020large}.

The spatial network models used here allow for complex social dependence through a kernel function, referred to as the \emph{social interaction function} or SIF. The SIF formally defines the relationship between two individuals based on spatial proximity. For example it has been shown that many social interaction patterns obey the Zipf law \cite{zipf2016human}, where individuals are more likely to interact with others close by rather than far away (a pattern that holds even for online interactions \cite{spiro2016persistence}). Here, we use this approach to model a network that represents combination of frequent interactions due to ongoing social ties, and contacts resulting from frequent incidental encounters (e.g., interactions with neighbors and community members).

We follow the protocol of \cite{butts2012geographical,hipp2013extrapolative} to simulate social network data that combines the actual distribution of residents in a city with a pre-specified SIF. We employ the model and data from \cite{hipp2013extrapolative} to produce large-scale social networks for 19 cities and counties in the United States -- providing a representation of major urban areas in the United States (see supplement).  Given these simulated networks, then implement an SIR-like framework to examine COVID-19 diffusion.  At each moment in time, each individual can be in a \emph{susceptible}, \emph{infected but not infectious}, \emph{infectious}, \emph{deceased}, or \emph{recovered} state.  The disease diffuses through the contact network, with currently infectious individuals infecting susceptible neighbors as a continous time Poisson process with a rate estimated from mortality data (see supplement); recovered or deceased individuals are not considered infectious for modeling purposes.  Upon infection, an individual's transitions between subsequent states (and into mortality or recovery) are governed by waiting time distributions based on epidemiological data as described in the supplemantary materials.  To begin each simulated trajectory, we randomly infect 25 individuals, with all others being considered susceptible.  Simulation proceeds until no infectious individuals remain.


From the simulated trajectory data, we produce several metrics to assess spatial heterogeneity in disease outcomes. First, we present infection curves for illustrative cities, showing the detailed progress of the infection and its difference from what an SIR model would posit. We also present chloropleth maps showing spatial variation in peak infection times, as well as the correlations between the infection trajectory within local areal units and the aggregate infection trajectory for the city as a whole.  While an SIR model would predict an absence of systematic variation in the infection curves or the peak infection day for different areal units in the same city, geographically realistic models show considerable disparities in infection progress from one neighborhood to another.  To quantify the degree of heterogeneity more broadly, we examine spatial variation in outcomes for each of our city networks. We show that large variations in peak infection days within across tracts are typical (often spanning weeks or even months), and that overall correlations of within-tract infection trajectories with the aggregate urban trajectory are generally modest (a substantial departure from what would be expected from an SIR model).  

In addition to these relatively abstract metrics, we also examine a simple measure of the potential load on the healthcare system in each city. Given the locations of each hospital in each city, we attribute infections to each hospital using a Voronoi tessellation (i.e., under the simple model that individuals are most likely to be taken to the nearest hospital if they become seriously ill).  Examination of the potential hospital demand over time shows substantial differences in load, with some hospitals severely impacted while others have few cases.  Finally, we consider the \emph{social exposure} of individuals to COVID-19, by computing the fraction of individuals with a personal contact who is respectively morbid or deceased.  Our model shows considerable differences in these metrics over time, revealing that the pandemic can appear very different to those ``on the ground'' -- evaluating its progress by its impact on their own personal contacts -- than what would be suggested by aggregate statistics. 



\section*{Data}

\subsection*{Spatial Network Data}

Networks are generated using population distributions from \cite{hipp2013extrapolative}, based on block-level Census data.  Hospital information are obtained from the Homeland Infrastructure Foundation-Level Data (HIFLD) database \cite{dhs2016homeland}. HIFLD is a an initiative that collects geospatial information on critical infrastructure across multiple levels of government. We employ the national-level hospital facility database, which contains locations of hospitals for the 50 US states, Washington D.C., US territories of Puerto Rico, Guam, American Samoa, Northern Mariana Islands, Palau, and Virgin Islands; underlying data are collated from various state departments or federal sources (e.g., Oak Ridge National Laboratory). We employ all hospitals within our 19 target cities, excluding facilities closed since 2019.  Latitude/longitude coordinates and capacity information were employed to create a spatial database that includes information on the number of beds in each hospital. 

The dates of the first confirmed case and all the death cases for King County, where Seattle is located, were obtained from The New York Times, based on reports from state and local health agencies \cite{nyt2020data}. The death rate was calculated based on population size of each county from the 2018 American Community Survey, and employed to calibrate the infection rate (the only free parameter in the models used here); details are provided in the supplemental materials. 

We ran 10 replicates of the COVID-19 diffusion process in each of our 19 cities, seeding with 25 randomly selected infections in each replicate and following the course of the diffusion until no infectious individuals remained.  Simulations were performed using a combination of custom scripts for the R statistical computing system \cite{rteam:sw:2020} and the statnet library \cite{butts:jss:2008a,butts:jss:2008b,handcock.et.al:jss:2008}.  Analyses were performed using R.

\section*{Results}

\subsection*{Smooth Aggregate Infection Trajectories Can Mask Local Outbreak Dynamics}

\begin{figure*}
\begin{multicols}{2}
    \includegraphics[width=\linewidth]{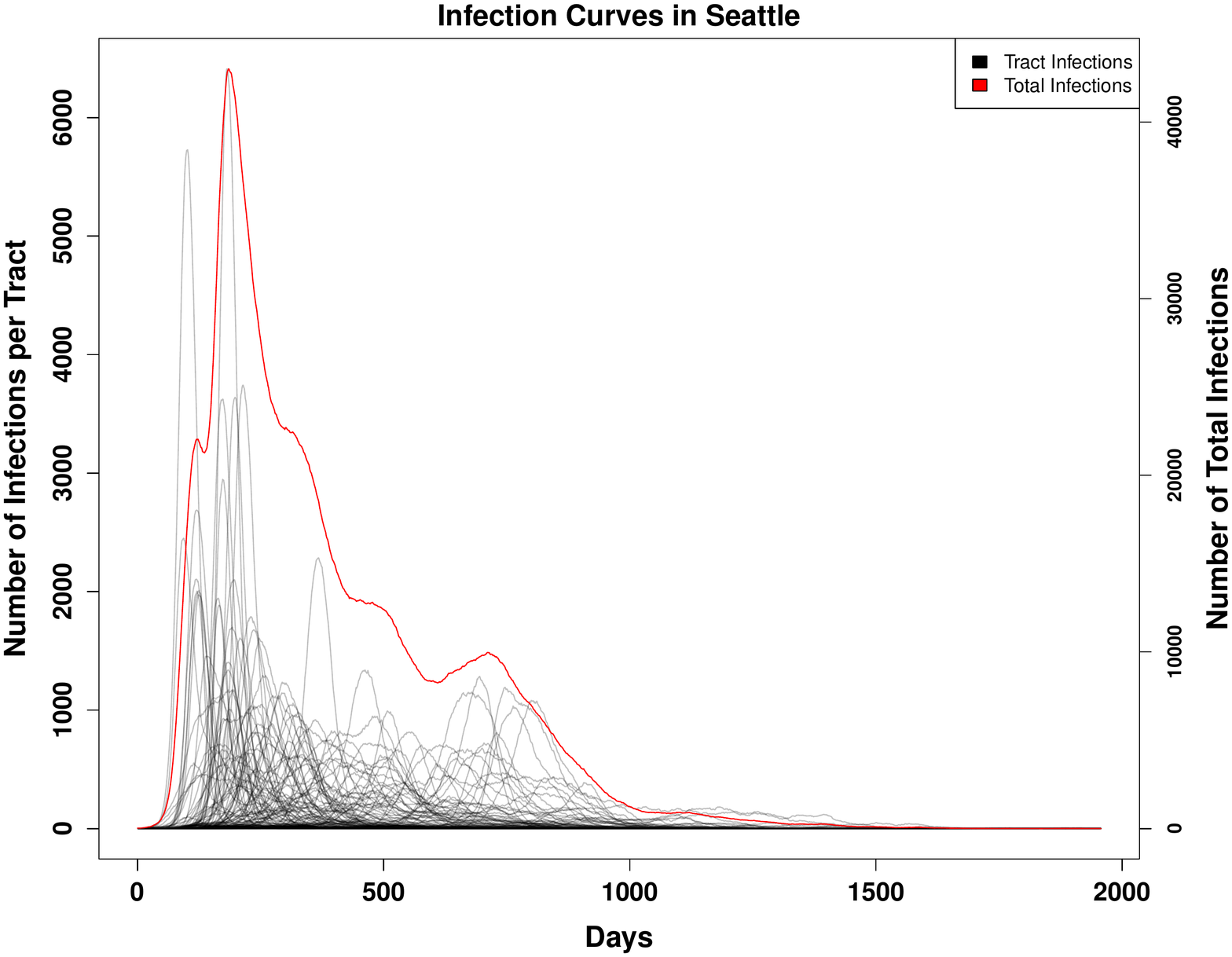} \par
    \includegraphics[width=\linewidth]{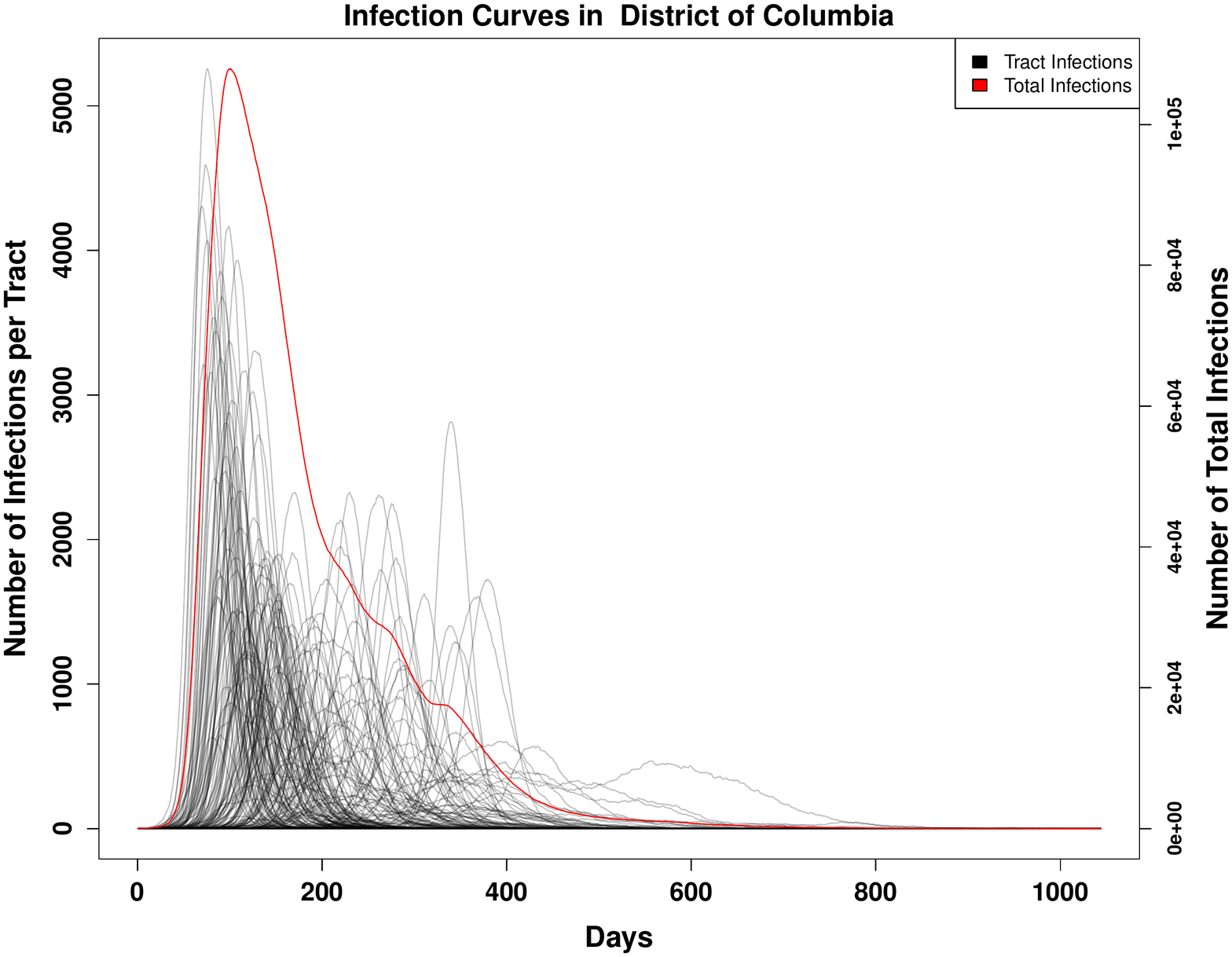} \par
\end{multicols}
\caption{(Left) Infection Curves for Seattle, WA. The Red line is the curve for the whole city, while the black lines are the the infection curves for each tract in the city. While the Red curve is relatively smooth, this smoothness hides a significant amount of heterogeneity in the timing of the infection curves for each census tract. (Right) Infection Curves for Washington D.C. As with Seattle, the city-level curve conceals considerable spatial variability in the infection's progress. \label{f_inf_curves}}
\end{figure*}

When taken over even moderately sized regions, aggregate infection curves can appear relatively smooth.  Although this suggests homogeneous mixing (as assumed e.g. by standard SIR models), appearances can be deceiving. Fig.~\ref{f_inf_curves} shows typical realizations of infection curves for two cities (Seattle, WA and Washington, DC), showing both the aggregate trajectory (red) and trajectories within individual Census tracts (black).  While the infection curves in both cases are relatively smooth, and suggestive of a fairly simple process involving a sharp early onset followed by an initially sharp but mildly slowing decline in infections, within-tract trajectories tell a different story.  Instead of one common curve, we see that tracts vary wildly in onset time and curve width, with some tracts showing peaks weeks or months after the initial aggregate spike has passed.

\begin{figure*}
\begin{multicols}{2}
    \includegraphics[width=\linewidth]{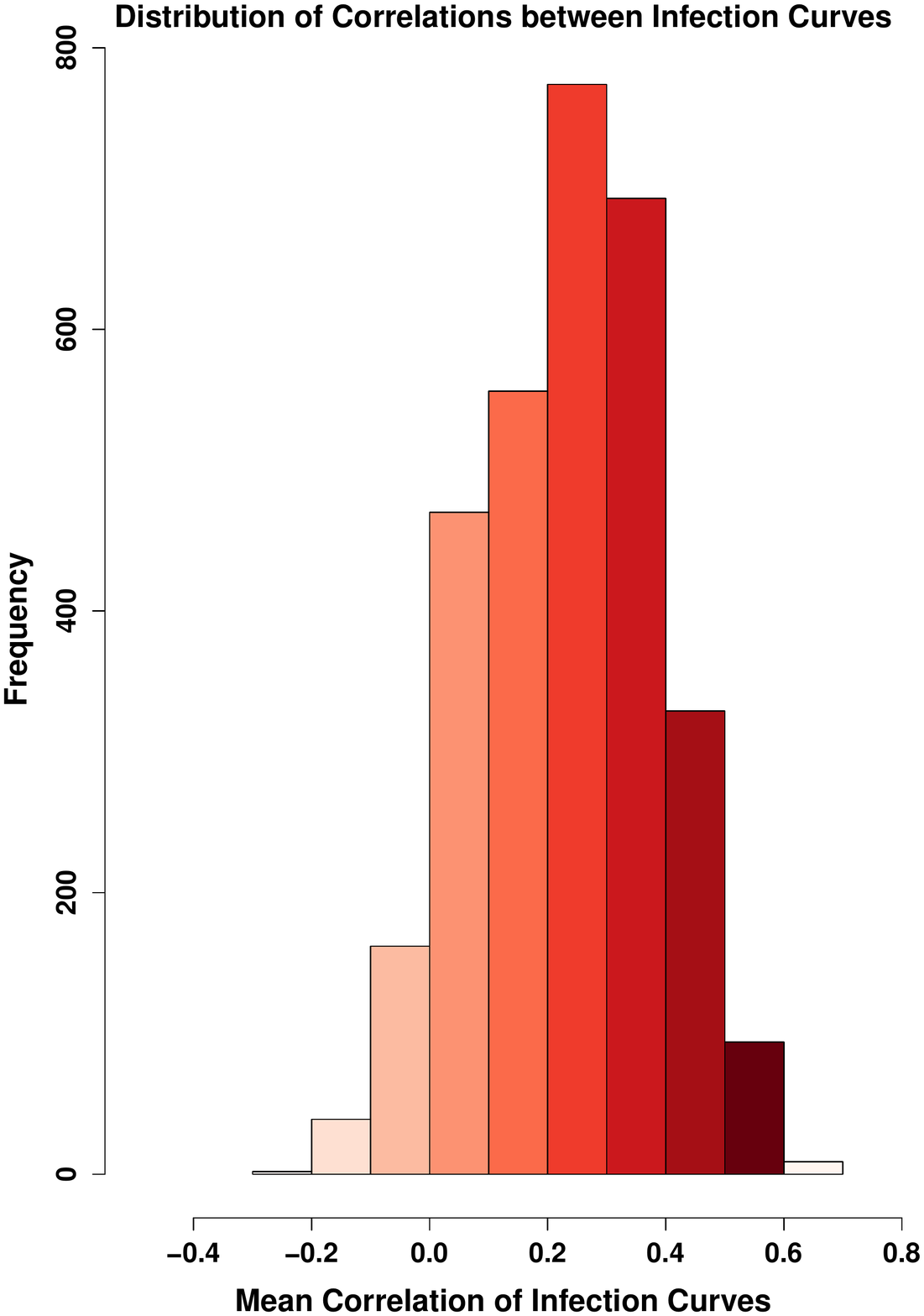} \par
    \includegraphics[width=\linewidth]{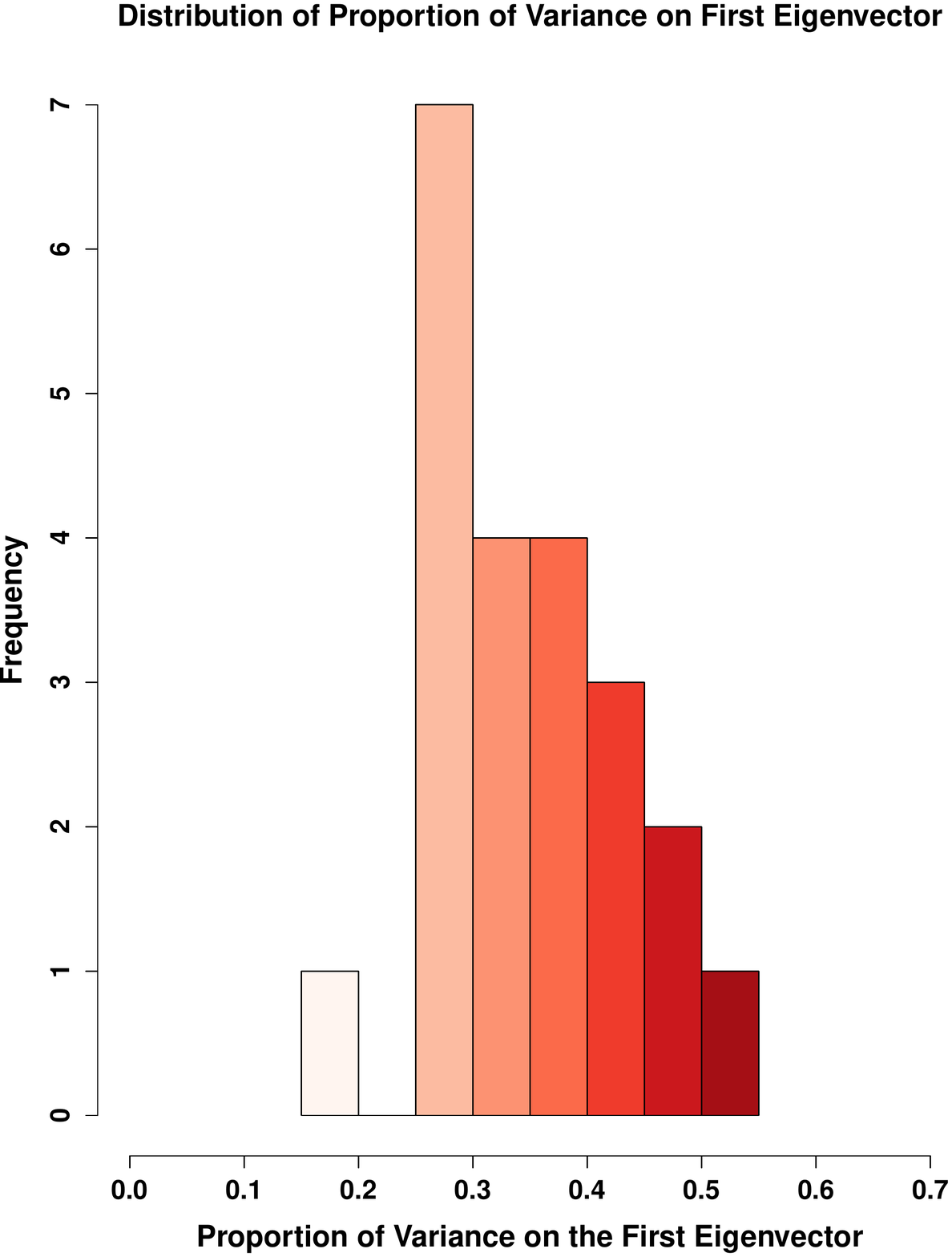} \par
\end{multicols}
\caption{ (Left) Histogram showing the mean pairwise correlation of infection curves for each tract within each city, across our entire sample.  The infection curve in any given tract is likely to have a correlation of only around 0.2 with any other tract in the city. This histogram includes a single datapoint for each tract in the sample.  (Right) Histogram of variance accounted for by the principal component of the standardized tract-level curve set.  None of the principal components account for more than 60\% of the variance, with most accounting for around only 35\% of the total variance. The datapoints included here include a single amount of variance explained for each city. \label{f_agg_cor}}
\end{figure*}

The cases of Fig.~\ref{f_inf_curves} are emblematic of a more systematic phenomenon: the progress of the infection within any given areal unit often has relatively little relationship to its progress in the city as a whole.  Fig~\ref{f_agg_cor} assesses this phenomenon over our entire sample, using two different consensus metrics.  First, we simply compute the correlation between the infection curves in each pair of tracts (assessed daily), taking the mean for each tract of its correlation with all other tracts within the city; if the progress of the infection were uniform across the city, the mean correlations would be large and positive.  Second, we provide a more direct assessment of the extent to which the set of infection curves can be summarized by a common pattern by taking the variance on the first principal component of the standardized infection curves.  As before, where different parts of the city experience similar patterns of growth and decline in infections, we expect the dimension of greatest shared variance to account for the overwhelming majority of variation in infection rates.  Contrary to these expectations, however, Fig.~\ref{f_agg_cor} shows that there is little coherence in tract-level infection patterns.  Mean correlations of local infection curves across tracts typically range from approximately 0 and 0.5, with a mean of approximately 0.2, indicating very little correspondence between infection timing in one track and that of another.  The principal component analysis tells a similar story: overall, we see that first component accounts for relatively little of the total variance in trajectories, with on average only around 35\% of variation in infection curves lying on the first principal component (and no observed case of the first component accounting for more than 60\% of the variance).  This confirms that \emph{local infection curves are consistently distinct}, with behavior that is only weakly related to infections in the city as a whole.  This is a substantially different scenario than what is commonly assumed in traditional SIR models.

\subsection*{Peak Infection Days Can Vary Substantially, Even Among Nearby Regions}

\begin{figure*}
\centering
\includegraphics[width=0.95\textwidth]{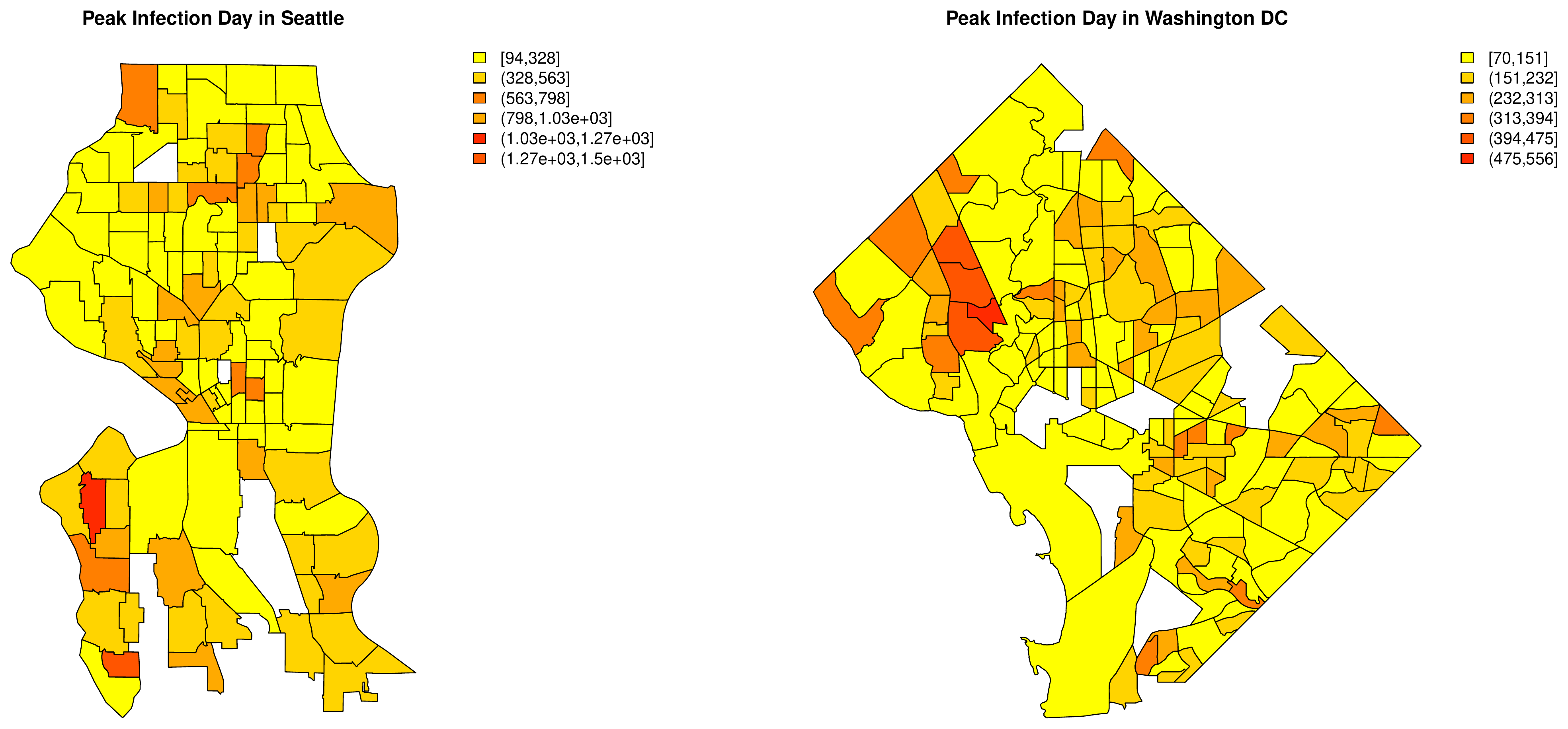} 
\caption{(Left) Chloropleth showing the peak day of infection in each tract in the city of Seattle. The map shows significant variability in when the infection peaks, with nearby regions sometimes having sharply different patterns. In parts of Western Seattle, the infection does not peak until almost a year past the first infections, while in the more northern parts of the city, the infection peaks much earlier. (Right) Times to peak infection for  Washington DC tracts. The northwestern part of the city has infections that are more delayed than in the central and northeastern parts of the city. Both of these maps show that there is a high degree of spatial heterogeneity present in the infection curves. \label{f_peak_map}}
\end{figure*}

These differences in local infection curves are a consequence of the unevenness of the ``social fabric'' that spans the city: while the disease can spread rapidly within regions of high local connectivity, it can easily become stalled upon reaching the boundaries of these regions. Further transmission requires that a successful infection event occur via a bridging tie, an event with a potentially long waiting time.  Such delays create potential opportunities for public health interventions (trace/isolate/treat strategies), but they can also create a false sense of security for those on the opposite side of bridge (who may incorrectly assume that their area was passed over by the infection).  Indeed, examining the time to peak infection across the cities of Seattle and Washington, D.C. (Fig.~\ref{f_peak_map}) shows that while peak times are visibly autocorrelated, tracts with different peak times frequently border each other. Residents on opposite sides of the divide may be exposed to very different local infection curves, making risk assessment difficult.

\begin{figure*}
	\includegraphics[width=\linewidth]{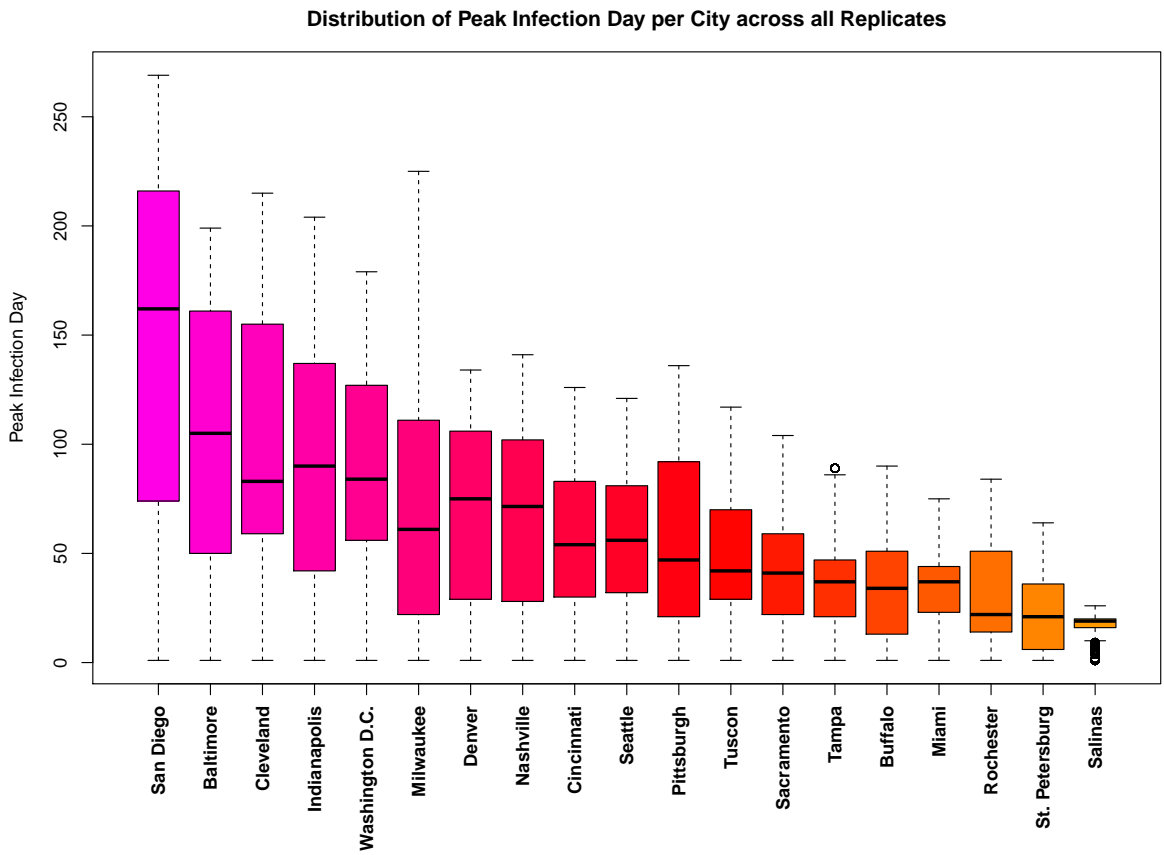}
\caption{Marginal distributions of days to peak infection by tract, across 10 simulated trajectories.  Though locales vary both in terms of overall median peak time and range of tract-level variation, large differences in peak time are nearly ubiquitous.  (Trajectory specific distributions shown in Fig.~S2.)  \label{f_marginal_peaks}}
\end{figure*}

The cases of Seattle and Washington, DC are not anomalous.  Looking across multiple trajectories over our entire sample, Fig.~\ref{f_marginal_peaks} shows consistently high variation in per-tract peak infection times for nearly all study communities.  (This variation is also seen within individual trajectories, as shown in supplemental figure~S2.)  Although peak times in some cities are concentrated within an interval of several days to a week, it is more common for peak times to vary by several months.  Such gaps are far from what would be expected under uniform local mixing.

\subsection*{Heterogeneous Impact Timing May Affect Hospital Load}

\begin{figure*}
\begin{multicols}{2}
    \includegraphics[width=\linewidth]{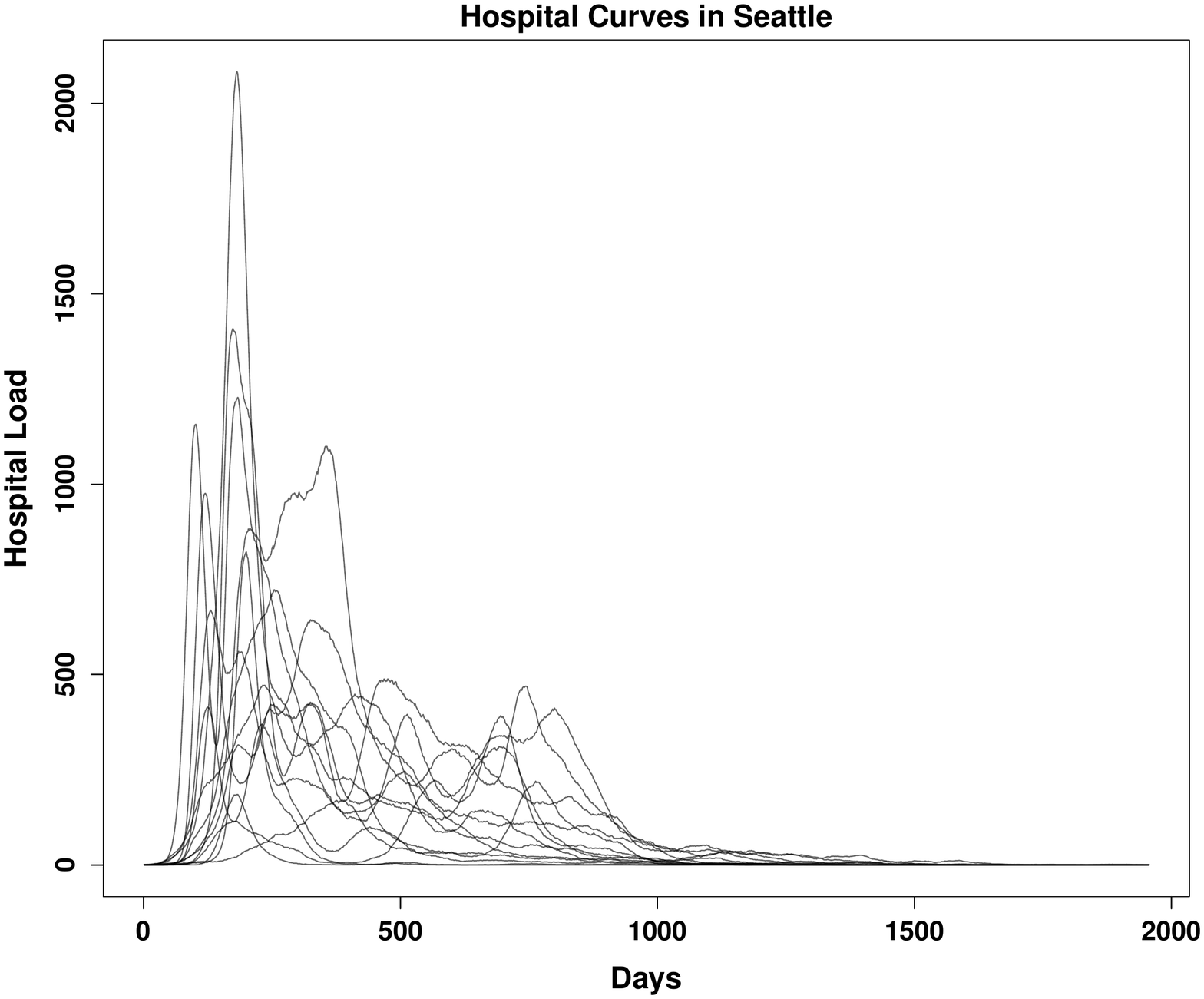} \par
    \includegraphics[width=\linewidth]{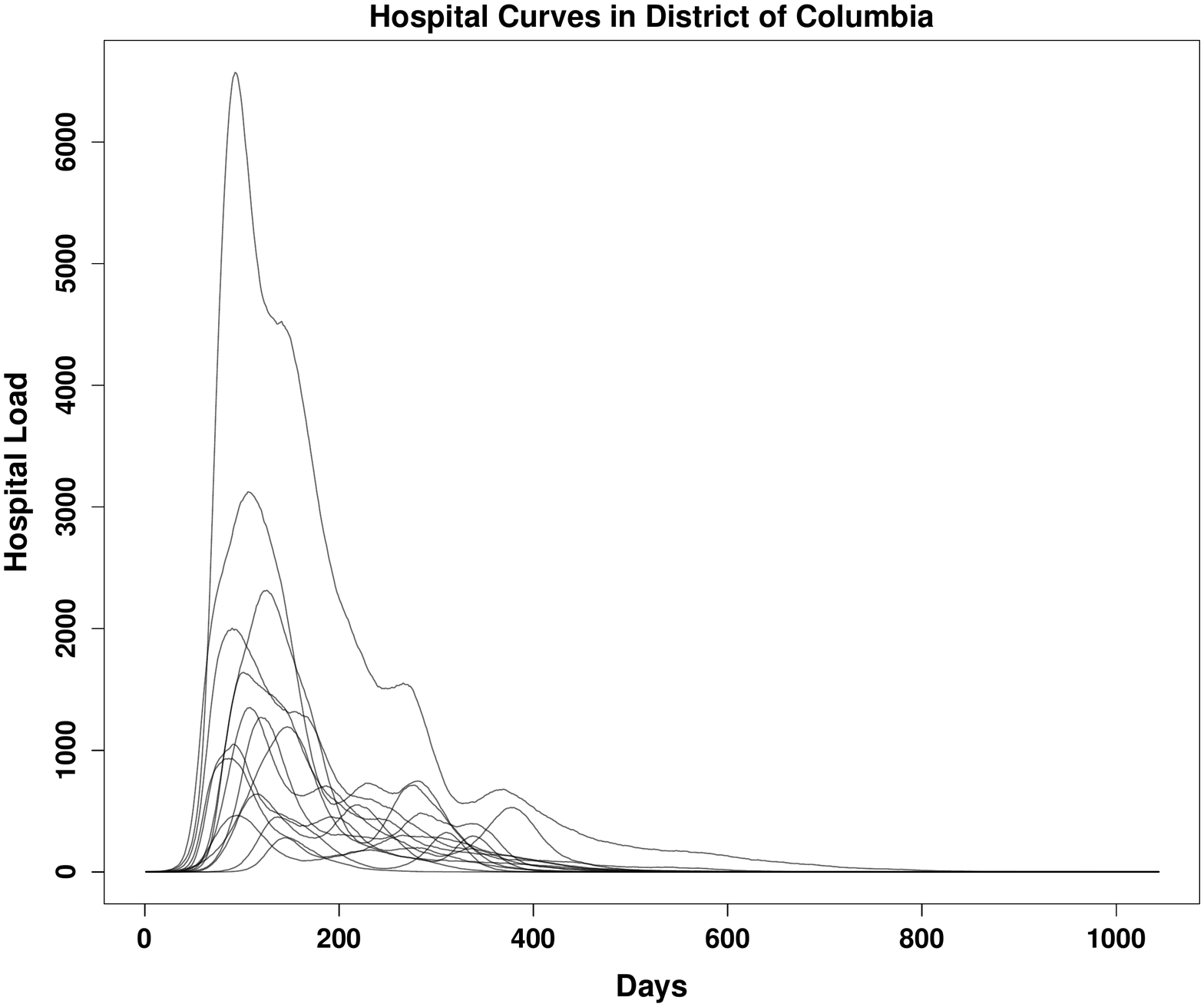} \par
\end{multicols}
\caption{(Left) Numbers of infections attributed to each hospital in the city of Seattle, with each curve representing a different hospital. Hospital peak demand times vary markedly, with some getting the majority of their hospitalizations before day 100, and others peaking almost a year into the pandemic. (Right)  Hospitalizations in Washington, DC.  As in Seattle, the each hospital has a unique demand trajectory, with some hospitals not getting their peak of infections until more than a year after the infection begins. \label{f_hosp_load}}
\end{figure*}

Variation in the timing of COVID-19 impacts across the urban landscape has potential ramifications for healthcare delivery, creating unequally distributed loads that overburden some providers while leaving others with excess resources.  To obtain a sense of how spatial heterogeneity in the infection curve could potentially impact hospitals, we employ a simple ``catchment'' model in which seriously ill patients are taken to the nearest hospital, subsequently recovering and/or dying as assumed throughout our modeling framework.  Based on prior estimates, we assume that 20\% of all infections are severe enough to require hospitalization\cite{covid2020severe}.  While hospitals draw from (and hence average across) areas that are larger than tracts, the heterogeneity shown in Fig.~\ref{f_inf_curves} suggests the potential for substantial differences in hospital load over time.  Indeed, our models suggest that such differences will occur.  Fig.~\ref{f_hosp_load} shows the number of patients arriving at each hospital in Seattle and Washington, DC (respectively) during a typical simulation trajectory.  While some hospitals do have demand curves that mirror the city's overall infection curve, others show very different patterns of demand.  In particular, some hospitals experience relatively little demand in the early months of the pandemic, only to be hit hard when infections in the city as a whole are winding down.

\begin{figure*}
\begin{multicols}{2}
    \includegraphics[width=\linewidth]{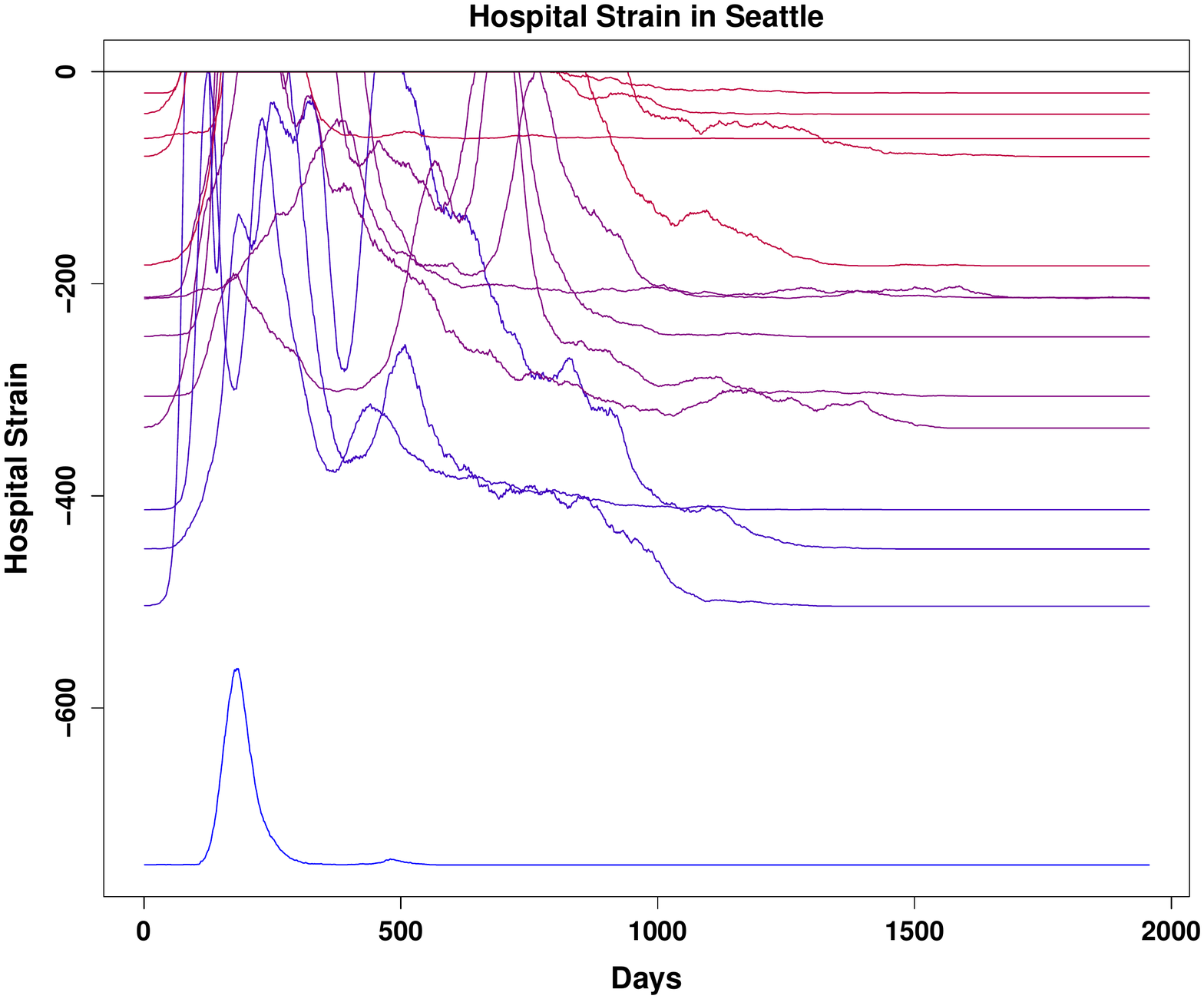} \par
    \includegraphics[width=\linewidth]{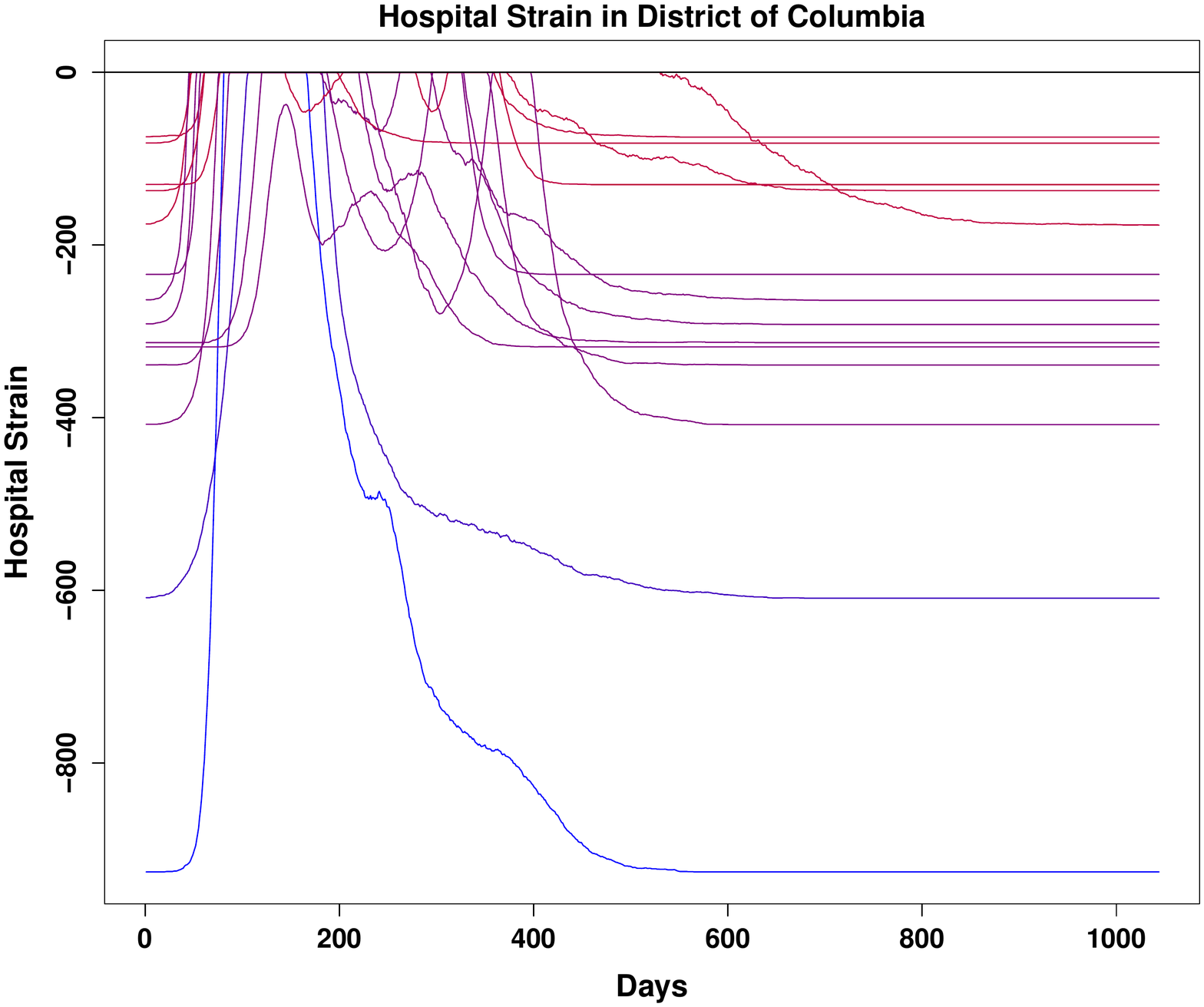} \par
\end{multicols}
\caption{(Left) Hospital strain in Seattle, WA. Values closer to zero indicate that hospitals are more strained and have fewer open beds, while lower values suggest more resources are available; color varies from blue (low average strain) to red (high average strain). Much like the number of infections, there is a high degree of heterogeneity present here, with hospitals freeing up resources at different points across the first year of the pandemic. (Right) Hospital strain for Washington D.C. Most hospitals get overwhelmed in the first 25 days of the pandemic, but then are able to recover at different times, usually within the second hundred days of the pandemic; some, however, are hit hard by a second wave, and others remain overwhelmed for several months.\label{f_hosp_strain}}
\end{figure*}

Just as hospital load varies, hospital capacities vary as well.  As a simple measure of strain on hospital resources, we consider the difference between the number of COVID-19 hospitalizations and the total capacity of the hospital (in beds), truncating at zero when demand outstrips supply.  (For ease of interpretation as a measure of strain, we take the difference such that higher values indicate fewer available beds.)  Using data on hospital locations and capacities, we show in Fig.~\ref{f_hosp_strain} strain on all hospitals in Seattle and Washington, D.C. (respectively) during a typical infection trajectory.  While some hospitals are hardest hit early on (as would be expected from the aggregate infection curve), others do not peak for several months.  Likewise, hospitals proximate to areas of the city with very different infection trajectories experience natural ``curve flattening,'' with a more distributed load, while those that happen to draw from positively correlated areas experience very sharp increases and declines in demand.  These conditions in some cases combine to keep hospitals well under capacity for the duration of the pandemic, while others are overloaded for long stretches of time.  These marked differences in strain for hospitals within the same city highlight the potentially complex consequences of heterogeneous diffusion for healthcare providers.

\begin{figure}[h!]
    \centering
    \includegraphics[width=0.5\textwidth]{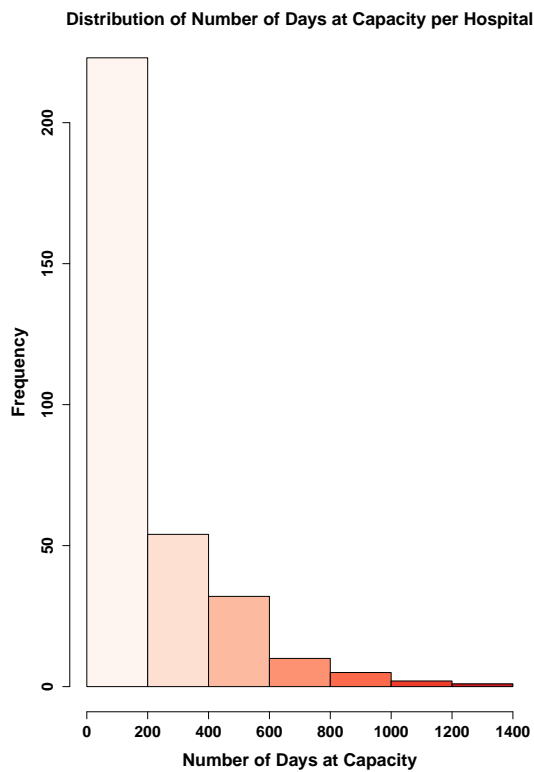}
    \caption{Marginal distribution of number of days without available beds, for all hospitals in our sample. While most hospitals will have only brief periods of overload, some will be at or over capacity for the entire pandemic, potentially several years.\label{f_hosp_overload}}
\end{figure}

Looking across cities, we see the same high-variability patterns as observed in Seattle and Washington.  In particular, we note that local variation in disease timing leads to a heavy-tailed distribution for the duration at which hospitals will be at capacity.  Fig.~\ref{f_hosp_overload} shows the marginal distribution of hospital overload periods (defined as total number of days at capacity during the pandemic), over the entire sample.  While the most common outcome is for hospitals to be stressed for a brief period (not always to the breaking point), a significant fraction of hospitals end up being overloaded for months - or even, in a small fraction of cases, nearly the whole duration of the pandemic. 

It should be reiterated that the hospital load model used here is extremely simplified, and that we are employing a no-mitigation scenario.  However, these results quite graphically demonstrate that the importance of curve-flattening interventions does not abate once geographical factors are taken into account.  On the other hand, these results suggest that differences in hospital load may be substantially more profound than would be anticipated from uniform mixing models, creating logistical challenges and possibly exacerbating existing differences in resource levels across hospitals.  At the same time, such heterogeneity implies that resource sharing and patient transfer arrangements could prove more effective as load-management strategies than would be suggested by spatially homogeneous models, as hospitals are predicted to vary considerably in the timing of patient demand.

\subsection*{Social Exposures to Morbidity and Mortality Vary by Location}

In addition to healthcare strain, the \emph{subjective experience} of the pandemic will potentially differ for individuals residing in different locations.  In particular, social exposures to outcomes such as morbidity or mortality may shape individuals' understandings of the risks posed by COVID-19, and their willingness to undertake protective actions to combat infection. Such exposures may furthermore act as stressors, with potential implications for physical and/or mental health.  As a simple measure of social exposure, we consider the question of whether a focal individual (ego) either has experienced a negative outcome themselves, or has at least one personal contact (alter) who has experienced the outcome in question.  (Given the highly salient nature of COVID-19 morbidity and mortality, we focus on the transition to first exposure rather than e.g. the total number of such exposures, as the first exposure is likely to have the greatest impact on ego's assessment of the potential severity of the disease.)   

\begin{figure*} [h!]
\begin{multicols}{2}
    \includegraphics[width=\linewidth]{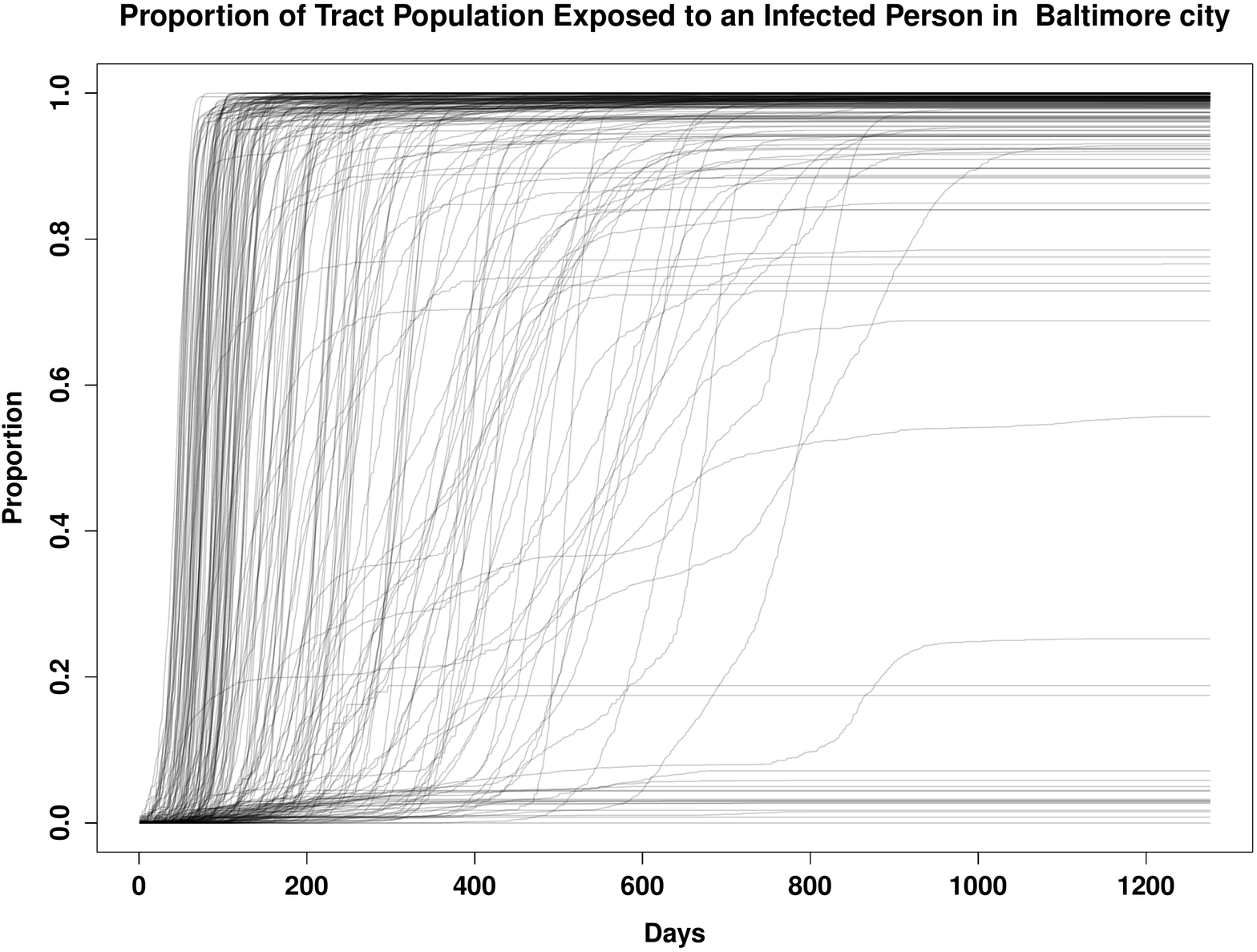} \par
    \includegraphics[width=\linewidth]{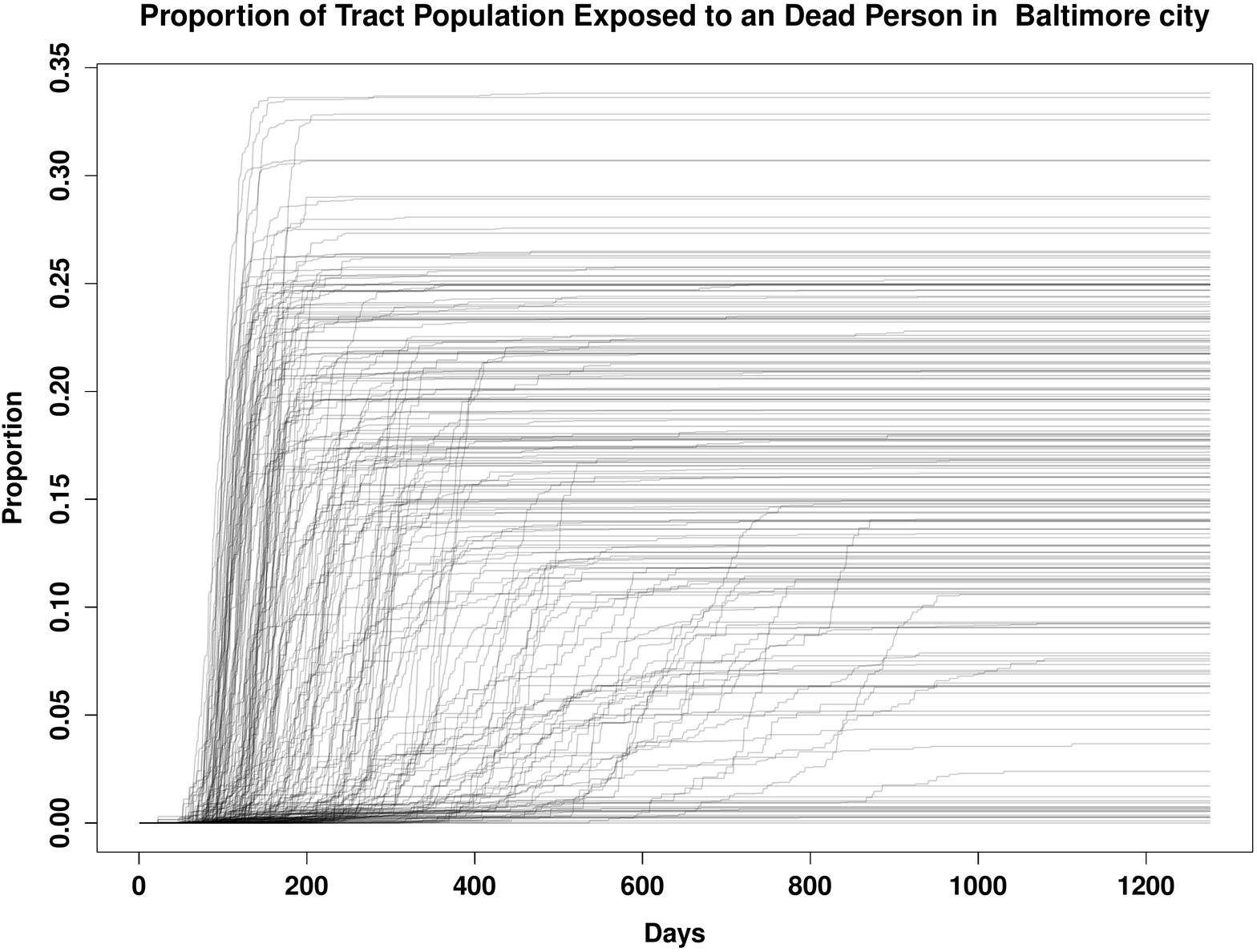} \par
\end{multicols}
\caption{(Left) Trajectories showing the fraction of people in each tract in Baltimore who have an infected person in their personal network across time. We see a large degree of spatial heterogeneity, as some tracts are more insulated from others in terms of social exposure. However, by the end of the pandemic, most people across all tracts have been exposed to someone who has had the disease. (Right) The fraction of persons in each tract who have an alter who died from COVID-19 in their personal network. On average, only around 40\% of people in any given tract know someone who died by the end of the pandemic, though this varies widely across tracts. \label{f_soc_exp}}
\end{figure*}

To examine how social exposure varies by location, we compute the fraction of individuals in each tract who are in socially exposed to (respectively) morbidity or mortality.  Fig.~\ref{f_soc_exp} shows these proportions for Baltimore, MD, over the course of the pandemic.  As with other outcomes examined here, we see considerable variation in timing, with many tracts seeing a rapid increase in exposure to infections, while others go for weeks or months with relatively few persons having a personal contact with the disease.  Another notable axis of variation is sharpness.  In many tracts, the fraction of individuals with at least one morbid contact transitions from near zero to near one within a matter of days, creating an extremely sharp social transition between the ``pre-exposure world'' (in which almost no one present knows someone with the illness) to a ``post-exposure world'' in which almost everyone knows someone with the illness).  By contrast, other tracts show a much more gradual increase (sometimes punctuated by jumps), as more and more individuals come to know someone with the disease.  In a few tracts that are never hit hard by the pandemic, few people ever have an infected alter; residents of these areas obviously have a very different experience than those of high-prevalence tracts.  These distinctions are even more stark for mortality, which takes longer to manifest and which does so much more unevenly.  Tracts vary greatly in the fraction of individuals who ultimately lose a personal contact to the disease, and in the repidity with which that fraction is reached.  In many cases, it may take a year or more for this quantity to be realized; until that point, many residents may be skeptical to the notion that the pandemic poses a great risk to them personally.

\begin{figure*} [h!]
\begin{multicols}{2}
    \includegraphics[width=\linewidth]{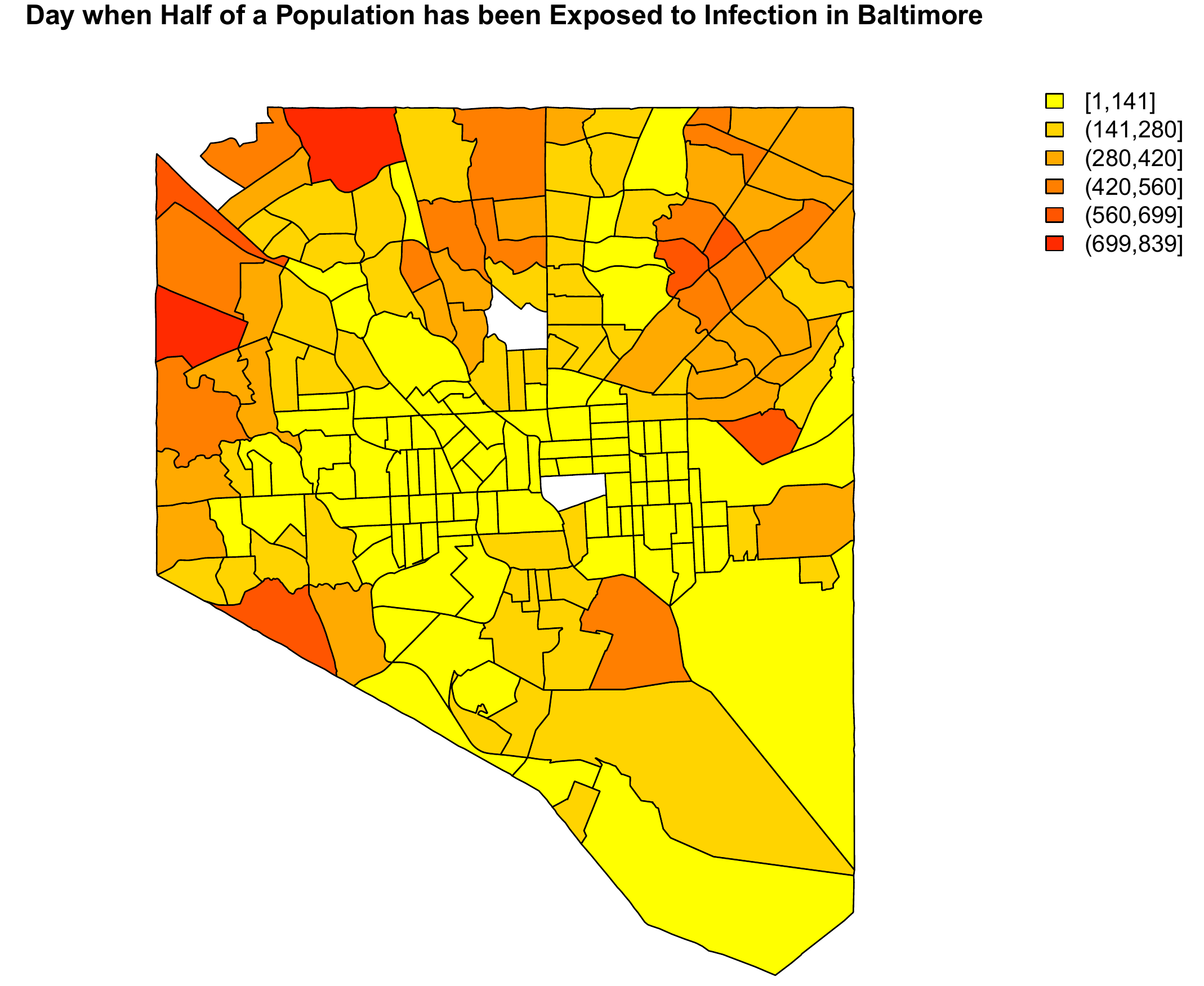} \par
    \includegraphics[width=\linewidth]{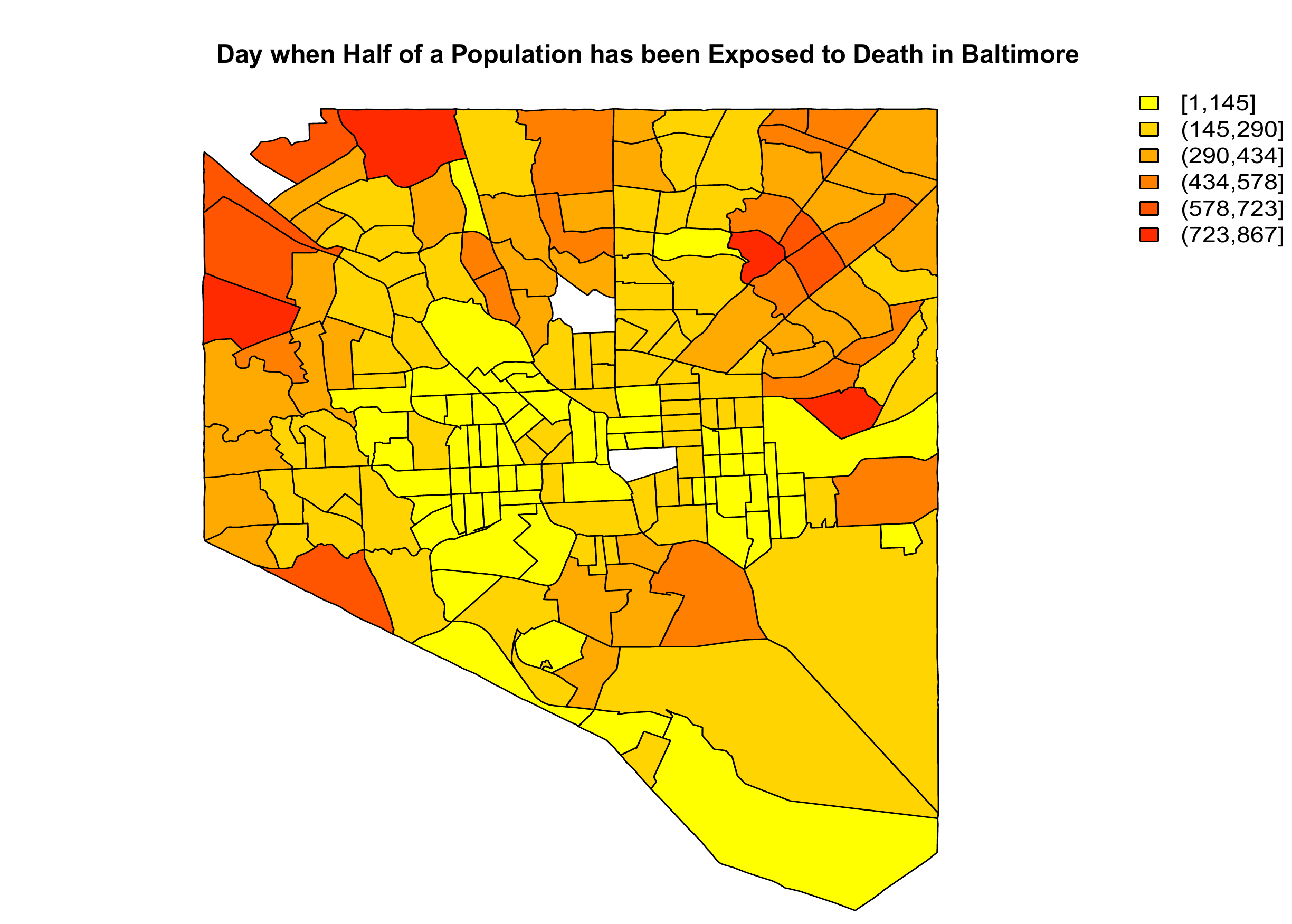} \par
\end{multicols}
\caption{(Left) Chloropleth showing the time for half of those in each tract to be socially exposed to COVID-19 morbidity in Baltimore, MD.   The central and southern parts of the city are exposed far sooner than the northwestern part of the city. (Right) Chloropleth showing the time for half of those in each tract to be socially exposed to COVID-19 mortality.  Central Baltimore is exposed to deaths in personal networks far sooner than the more outlying areas of the city. \label{f_exp_map}}
\end{figure*}

By way of assessing the milieu within each tract, it is useful to consider the ``cross-over'' point at which at least half of the residents of a given tract have been socially exposed to either COVID-19 morbidity or mortality.  Fig.~\ref{f_exp_map} maps these values for Baltimore, MD.  It is immediately apparent that social exposures are more strongly spatially autocorrelated than other outcomes considered here, due to the presence of long-range ties within individuals' personal networks.  Even so, however, we see strong spatial differentiation, with residents in the urban core being exposed to both morbidity and mortality much more quickly than those on the periphery.  This suggests that the social experience of the pandemic will be quite different from those in city centers than in more outlying areas, with the latter taking far longer to be exposed to serious consequences of COVID-19.  This may manifest in differences in willingness to adopt protective actions, with those in the urban core being more highly motivated to take action (and perhaps resistant to rhetoric downplaying the severity of the disease) than those on the outskirts of the city.

\section*{Discussion and Conclusion}

Our simulation results all underscore the potential effects of local spatial heterogeneity on disease spread. The spatial heterogeneity heterogeneity driving these results occurs on a very small scale (i.e., Census blocks), operating well below the level of the city a whole. As the infection spreads, relatively small differences in local network connectivity and the prevalence of bridging ties driven by uneven population distribution can lead to substantial differences in infection timing and severity, leading different areas in each city to have a vastly different experience of the pandemic.  Resources will be utilized differently in different areas, some areas will have the bulk of their infections far later than others, and the subjective experience of a given individual regarding the pandemic threat may differ substantially from someone in a different area. These behaviors are in striking contrast to what is assumed by models based on the assumption of spatially homogeneous mixing, which posit uniform progress of the infection within local areas.

As noted at the outset, our model is based on a no-mitigation scenario, and is not intended to capture the impact of social distancing.  While distancing measures by definition limit transmission rates - and will hence slow diffusion - contacts occurring through spatially correlated networks like those modeled here are still likely to show patterns of heterogeneity like those described.  One notable observation from our simulations is the long outbreak delay that some census tracts experience, even in the absence of social distancing. This would suggest that relaxation of mitigation measures leading to a resumption of ``normal'' diffusion may initially appear to have few negative effects, only to lead to deadly outbreaks weeks or months later.  Public health messaging may need to stress that apparent lulls in disease progress are not necessarily indicators that the threat has subsided, and that areas ``passed over'' by past outbreaks could be impacted at any time.


Finally, we stress that conventional diffusion models using locally homogeneous mixing have been of considerable value in both pandemic planning and scenario evaluation.  Our findings should not be taken as an argument against the use of such models.  However, the observation that incorporating geographical heterogeneity in contact rates leads to radically different local behavior would seem to suggest that there is value in including such effects in models intended to capture outcomes at the city or county level.  Since these are the scales on which decisions regarding infrastructure management, healthcare logistics, and other policies are often made, improved geographical realism could potentially have a substantial impact on our ability to reduce lives lost to the COVID-19 pandemic.

\section*{Acknowledgments}
This research was supported by NSF awards IIS-1939237 and SES-1826589 to C.T.B, and by the UCI Seed Grants program.

\bibliography{covid}
\bibliographystyle{Science}

\clearpage

\setcounter{table}{0}
\renewcommand{\thetable}{S\arabic{table}}
\setcounter{figure}{0}
\renewcommand{\thefigure}{S\arabic{figure}}


\section*{Supplement to ``Spatial Heterogeneity Can Lead to Substantial Local Variations in COVID-19 Timing and Severity''}


In this supplement, we expand on several of the technical sections that are breifly discussed in the main text. We specifically discuss in more detail the construction of the networks used, and the estimation of the parameters for the simulation study.


\subsection*{Introduction}

In this supplement, we go into more depth on Spatial Interaction Functions, Spatial Bernoulli Models, the setup and parameterizations of our simulations, and the paramter estimation problems that we used for this paper. We focus specifically on the more technical aspects of each of these components, showing how they have been formally specified and parameterized.

\subsubsection*{Spatial Interaction Function}

A Spatial Interaction Function (SIF) describes the marginal probability of a tie between any two nodes, given the distance between those nodes, represented  $\mathcal{F}(\mathcal{D}_{i,j},\theta)$. In this representation, $\mathcal{D}_{i,j}$ is the distance between the nodes, and $\theta$ are the parameters for the function. Prior literature shows that spatial interaction functions tend to be of the power law or attenuated power law form \cite{butts2011spatial}. Thus, we can represent the SIF as  $\mathcal{F}(\mathcal{D}_{i,j},\theta) = \frac{p_b}{(1 + \alpha * \mathcal{D}_{i,j})^\gamma}$. Here, $p_b$ represents the base tie probability, which can be thought of as the probability of a tie at distance 0. $\alpha$ is a scaling parameter that determines the speed at which the probability drops towards zero. $\gamma$ is the parameter that determines the weight of the tail. 

We draw on two SIFs in this paper, using models for social interactions and face to face interactions employed in prior studies \cite{butts2012geographical,hipp2013extrapolative}. The social interaction SIF declines with a $\gamma$ of 2.788, while the face-to-face SIF declines with $\gamma$ of 6.437. The paramters for the social interaction SIF are $p_b = 0.533$, $\alpha = 0.032$, $\gamma = 2.788$, and the parameters for the face-to-face SIF are $p_b = 0.859$,$\alpha = 0.035$, $\gamma=6.437$ \cite{hipp2013extrapolative}.


\subsection*{Spatial Bernoulli Models}

 Bernoulli Models are a class of random graph models that leverage the concept of a Bernoulli graph, or a graph in which the probability associated with each edge is a Bernoulli trial. In a spatial Bernoulli graph, tie probabilities are determined by a Spatial Interaction Function, applied to the pairwise distances between individuals within some space (here, geographically determined using Census data). Spatial Bernoulli models are highly scalable due to the conditional independence of edges, but allow for extremely complex structure due to the heterogeneity in edge probabilities induced by the SIF; likewise, they naturally produce properties such as local cohesion and degree heterogeneity observed in many types of social networks \cite{butts2012geographical}.  Formally, we can specify a Spatial Bernoulli Model by $Pr(Y_{ij}=1)=\mathcal{F}(\mathcal{D}_{i,j},\theta)$, where $Y_{ij}$ is each dyad, and $\mathcal{F}(\mathcal{D}_{i,j},\theta)$ is a Spatial Interaction Function with inputs as distance $\mathcal{D}_{i,j}$, and parameters $\theta$.

\subsection*{Network Simulations}

To simulate diffusion of COVID-19, we require a contact network. Here, we employ the above-described spatial Bernoulli graphs, with node locations for each of our 19 study locations drawn based on block-level Census data (including clustering within households, an important factor in disease diffusion). We follow the protocols described in \cite{almquist2012point,butts2012geographical} to generate node positions, specifically using the quasirandom (Halton) placement algorithm. Node placement begins with the households in each census block, using Census data from \cite{hipp2013extrapolative}. The quasirandom placement algorithm uses a Halton sequence to place households in space within the areal unit in which they reside. If any two households are placed within a critical radius of each other, then the algorithm ``stacks'' the households on top of each other by introducing artificial elevation (simulating e.g. a multistory apartment building).  Once all households are placed, individuals within households are placed at jittered locations about the household centroid.  (Individuals not otherwise attached to households are treated as households of size 1.) 

Given an assignment of individuals to spatial locations, we simulate spatial Bernoulli graphs using the models specified above.  We generate two networks for each city, one with the social interaction SIF, and the other with the face-to-face interaction SIF. To form a network of potential high-risk contacts, we then merge these networks (which share the same node set) by taking their union, leading to a network in which two individuals are tied if they either have an ongoing social relationship or would be likely to have extensive face-to-face interactions for other reasons (e.g., interacting with neighbors).  This process is performed for each city in our sample.

\subsection*{List of Cities}

Table S1 lists the cities that we use for our simulations. These data are drawn from \cite{hipp2013extrapolative}.

\begin{table}[h!]
\centering
\caption{List of study communities. \label{tb:citylist}}
\begin{tabular}{rl}
  \hline
 & City/County\\ 
  \hline
  1 & Buffalo \\ 
  2 & Baltimore \\ 
  3 & Cincinnati \\
  4 & Cleveland \\
  5 & Denver \\
  6 & Indianapolis \\ 
  7 & Miami \\
  8 & Milwaukee \\
  9 & Nashville \\
  10 & Pittsburgh \\ 
  11 & Rochester \\ 
  12 & Sacramento \\
  13 & Salinas \\ 
  14 & San Diego City \\ 
  15 & Seattle \\ 
  16 & St. Petersburg \\ 
  17 & Tampa \\ 
  18 & Tuscon \\ 
  19 & Washington DC \\ 
   \hline
\end{tabular}
\end{table}

\subsection*{Disease Simulations}


%

We conduct a series of simulations to examine the spread of COVID-19 across city-sized networks. These simulations use a simple continuous-time network diffusion process, the general description of which are described in the main text. The input for the diffusion simulation is a network and a vector of initial disease states (\emph{susceptible}, \emph{latent} (infected but not yet infectious), \emph{infectious}, \emph{recovered}, and \emph{deceased}), and the output is detailed history of the diffusion process up to the point at which a steady state is obtained (i.e., no infectious individuals remain).  Infection occurs via the network, with currently infectious individuals infecting susceptible alters as Poisson events with a fixed rate.  The transitions between latent and infectious, and infectious and either recovery or mortality are governed by gamma distributions estimated from epidemiological data. Table \ref{tb:gamma_params} shows the estimated shape and scale parameters for the gamma distributions employed here. The parameters for waiting time to infectiousness are directly available in the Appendix of \cite{lauer2020incubation}, while those for the recovery and death are estimated by matching the mean and standard deviation of durations reported in the literature \cite{verity2020estimates}.  Selection into death versus recovery was made via a Bernoulli trial drawn at time of infection (thereby determining which waiting time distribution was used), with the estimated mortality probabiliy being 0.0215.
 
\begin{table}[ht]
\centering
\caption{Shape and Scale parameters for Gamma distributions for durations (unit: day). \label{tb:gamma_params}}
\begin{tabular}{llll}
  \hline
 & Death & Recovery & Infectious \\ 
  \hline
Shape & 4.566 & 5.834 & 5.807 \\ 
Rate  & 0.251 & 0.219 & 1.055 \\
Scale (i.e., 1/Rate) & 3.984 & 4.566 & 0.948 \\ 
   \hline
\end{tabular}
\end{table}



\subsection*{Infection Rate Parameter Estimation}

To determine the infection rate (the only free parameter for the network models used in our simulations), we simulate the diffusion of virus in Seattle and fit it to the over-time death rate of the King County, WA before the first shelter-in-place order went into effect on March 23, 2020. (We limit our data to this time period because our simulation employs a no-mitigation scenario.)  A grid search strategy was employed to determine the expected days to transmission (which is the inverse of infection rate), and the number of days between the existence of the first infected cases and the first confirmed cases (aka the time lag, a nuisance parameter that is relevant only for estimation of the infection rate). The time lag is treated as an integer and the expected days to transmission as a continuous variable. For each lag/rate pair, we randomly take 5 draws from the expected infection waiting time distribution, add them to the lag time (i.e. the introduction of the true patient zero for the initial outbreak), and simulate 50 realizations of the diffusion process (redrawing the network each time). The diffusion rate parameter was selected based on minimizing the mean squared error between the simulated death rate and the observed number of deaths over the selected period. The first round of grid-search divided the expected days of search into 100 intervals, from (1,2) to (100,101), with days of lag ranging from 1 to 100 days. The second round of grid-search, based on the performance of the first round, divided the expected days of search into 240 intervals, from (41.00,41.25) to (100.75,101.00), with days of lag ranging from 1 to 60 days. The grid-search suggests that the expected days to transmission is 78.375 (78.25,78.50) days (Fig S1); that is, in a hypothetical scenario in which a single infective ego remained indefinitely in the infective state, and a single alter remained otherwise susceptible, the average waiting time for ego to infect alter would be approximately 80 days.  While this may at first blush appear to be a long delay, it should be borne in mind that this embodies the reality that no individual is likely to infect any \emph{given} alter within a short period (since, indeed, ego and alter may not happen to interact within a narrow window).  With many alters, however, the chance of passing on the disease is quite high.  Likewise, we note that the thought experiment above should not be taken to imply that actors remain infectious for such an extended period of time; per the above-cited epidemiological data, individuals typically remain infectious for roughly 18-33 days (though variation outside this range does occur, as captured by the above gamma distributions).  When both delay times are considered, the net probability of infecting any given alter prior to recovering is approximately 27\%.

Using the above, we can calculate the corresponding basic reproductive number ($R_0$):
\begin{equation}
R_0=d \alpha \tau
\end{equation}
where $d$ is the mean degree of an individual in the network; $\alpha$ is the infection rate (the inverse of expected days of transmission); $\tau$ is the time spent in the infectious state (here in days). For each simulated Seattle contact network, we calculate the degree for every individual. The time in the infectious statae was obtained by simulating gamma distributions for days of incubation, recovery, and death, and randomly permuting the distribution for 10 times for each simulated SIF network. Taking the mean of $R_0$ for each individual, the corresponding basic reproductive number in the diffusion simulation model is 1.7.  

\subsection*{Simulation Replicates}

To supplement the results on the variation in the peak infection time given in the main text, we ran a series of simulation replicates. Figure 3 in the main text shows the data from the figure below aggregated across all replicates. In the supplemental figure S2, we break out the peak infection days in each city, by replicate. These data show that the significant variation in Figure 3 is not due to the number of replicates that were run, but instead due to the intrinsic variation that is present (due to spatial heterogeneity).

\begin{figure*}
\centering
    \begin{subfigure}[b]{0.495\textwidth}
        \includegraphics[width=\textwidth]{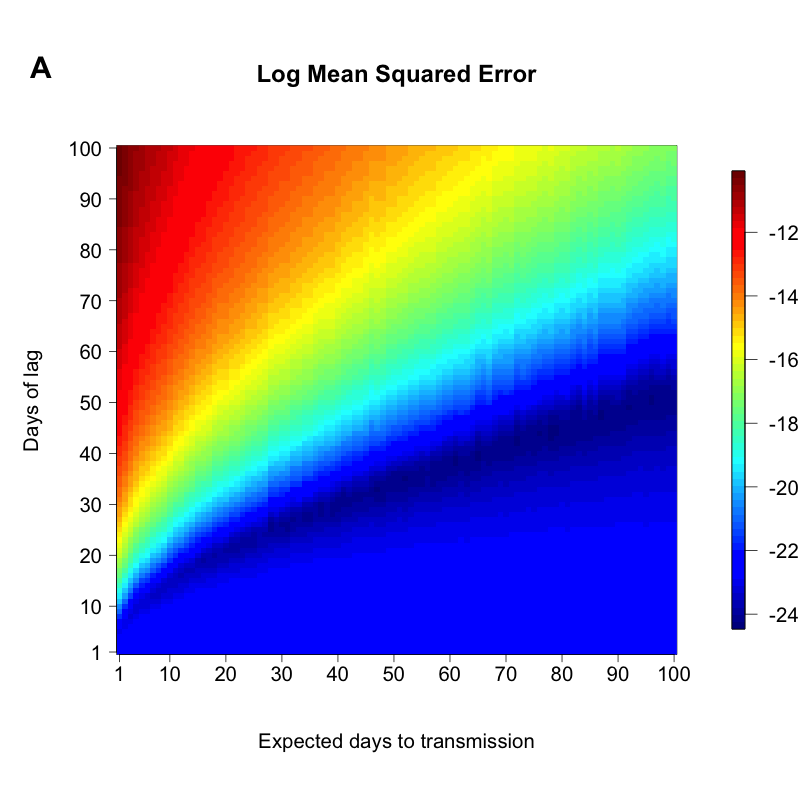}
    \end{subfigure}
    \begin{subfigure}[b]{0.495\textwidth}
        \includegraphics[width=\textwidth]{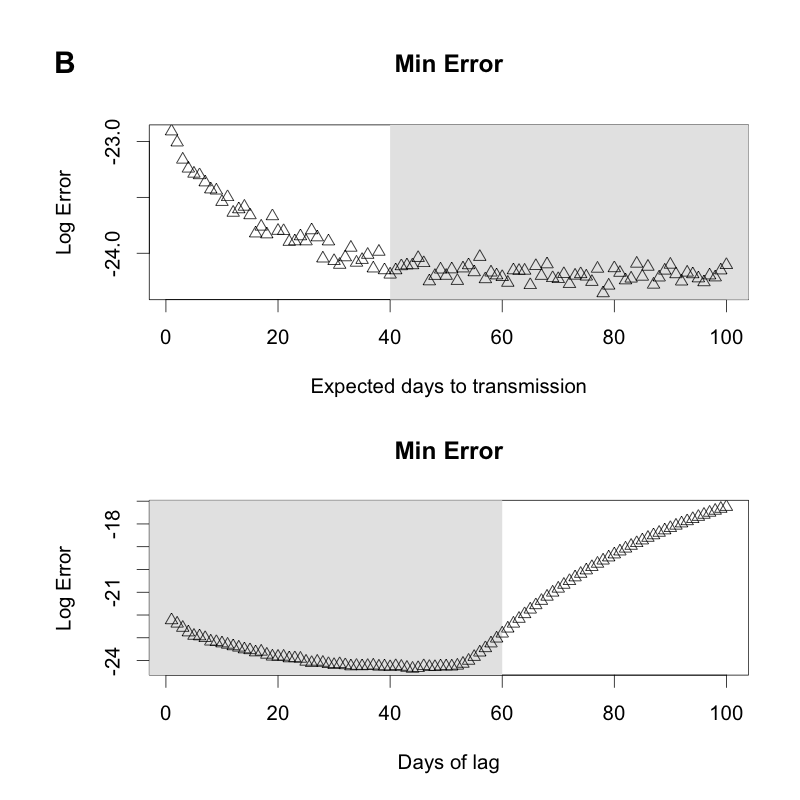}
    \end{subfigure}
    \begin{subfigure}[b]{0.495\textwidth}
        \includegraphics[width=\textwidth]{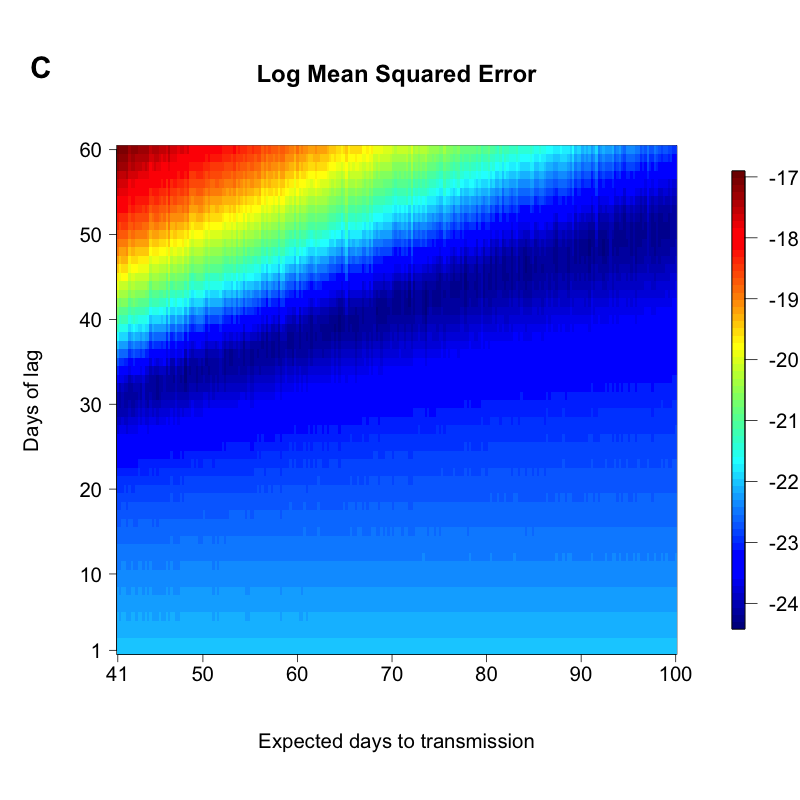}
    \end{subfigure}
    \begin{subfigure}[b]{0.495\textwidth}
        \includegraphics[width=\textwidth]{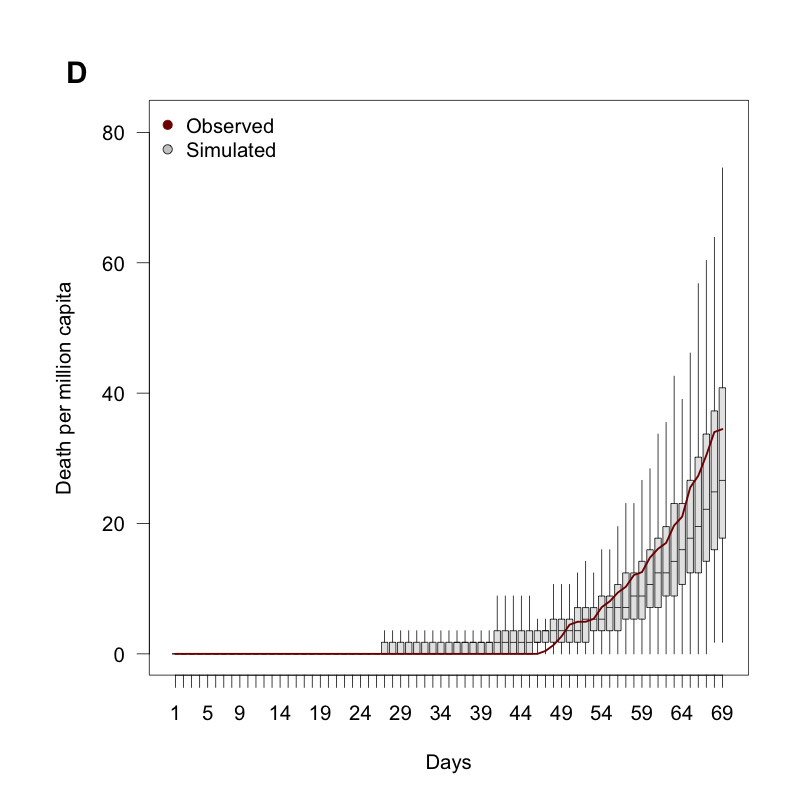}
    \end{subfigure}
    \caption{\textbf{Grid-search for infection rate.} The mean squared error (in logarithm form) for simulations with combinations of days of lag and expected days to transmission, for the first round (\textbf{A}) and the second round (\textbf{C}). The cross-sectional analysis of the minimum error for each days of lag and expected days to transmission (\textbf{B}) suggests the interval of both variables for the second round of search (areas colored in gray). Curves of death rate based on the best-fit parameter: 45 days of lag, 78.375 expected days to transmission (\textbf{D}). \label{fig:gridsearch}}
\end{figure*}

\begin{figure*}
\centering
    \begin{subfigure}[b]{0.495\textwidth}
        \includegraphics[width=\textwidth]{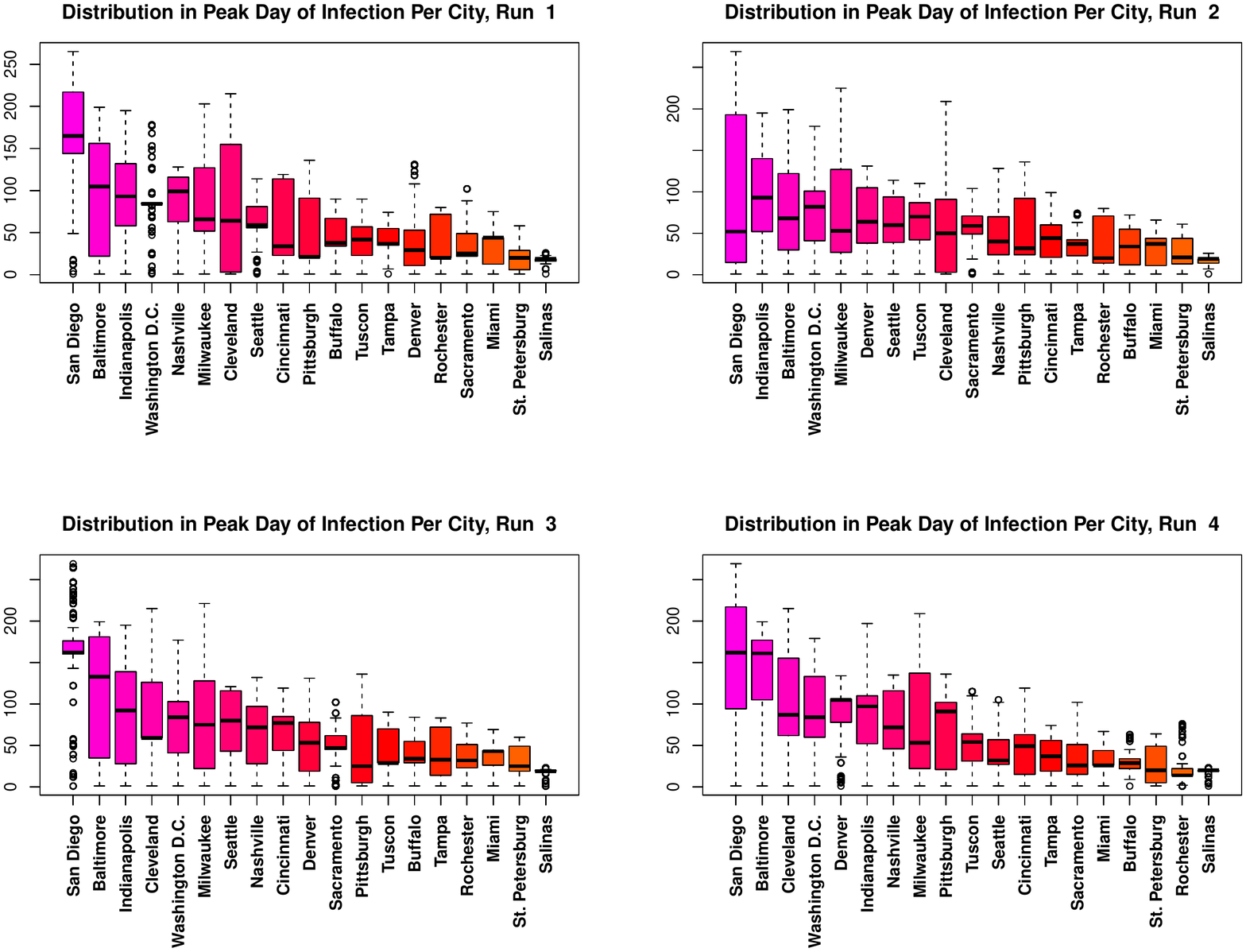}
    \end{subfigure}
    \begin{subfigure}[b]{0.495\textwidth}
        \includegraphics[width=\textwidth]{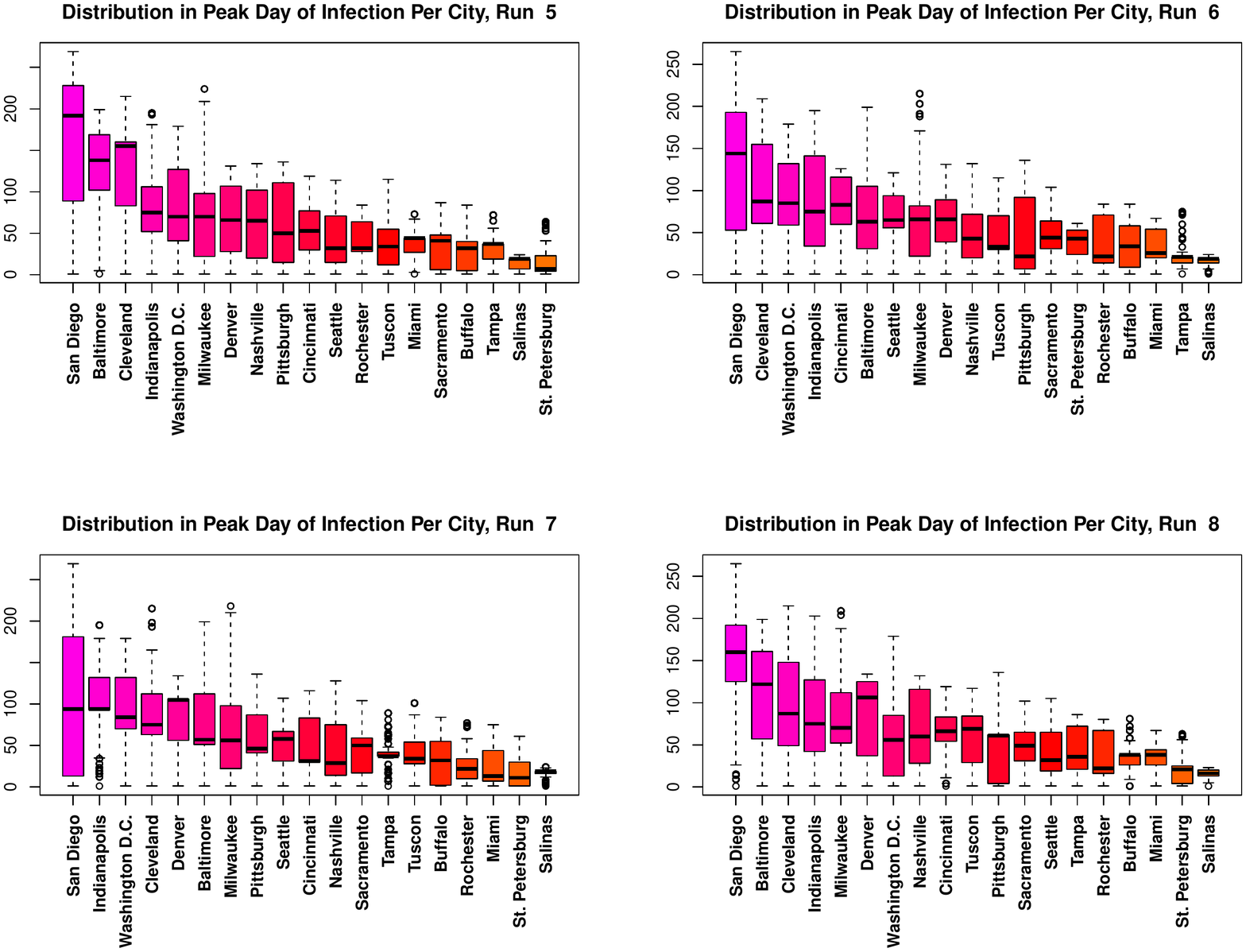}
    \end{subfigure}
    \begin{subfigure}[b]{0.495\textwidth}
        \includegraphics[width=\textwidth]{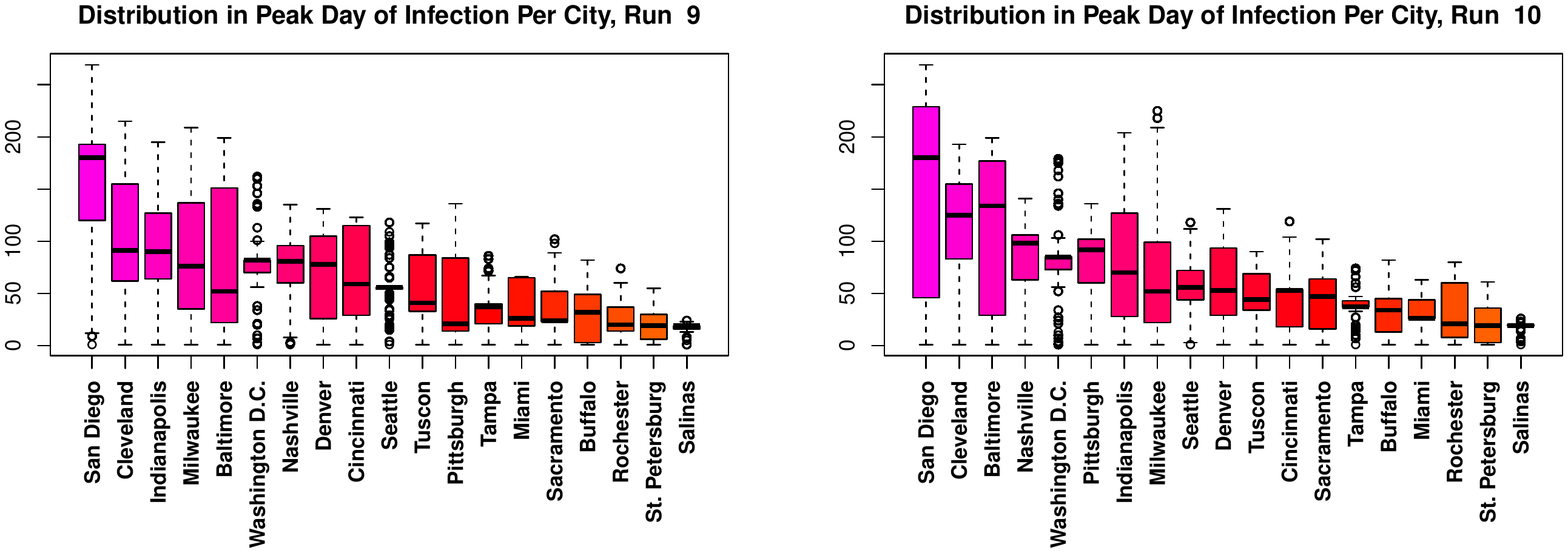}
    \end{subfigure}
\caption{Boxplot showing the peak infection days across 10 replicates for each city in the sample. There is a large degree of heterogeneity within each city, showing that the day that the infection peaks for any given tract is not uniform at all. Within any given city, there is a consistently high amount of variance in the peak infection day. In other words, the variance that we show here is a property of the spread of the disease, rather than the number of simulation replicates. \label{f_peakreps}}
\end{figure*}


\end{document}